\documentclass[reprint,nofootinbib,superscriptaddress,onecolumn]{revtex4-2}

\usepackage{amsmath,amssymb}
\usepackage{orcidlink}
\usepackage{float}
\usepackage{bm}
\usepackage{hyperref,appendix}
\usepackage[T1]{fontenc}

\usepackage[ISO]{diffcoeff}[=v4]

\newcommand{\pdv}[3][1]{\diffp[#1]{#2}{#3}}
\newcommand{\fdv}[3][1]{\diff.delta.[#1]{#2}{#3}}
\diffdef{}{long-var-wrap = dv}

\newcommand{\dd}{\mathrm{d}}
\newcommand{\Eff}{\mathcal F}

\newcommand{\A}{\mathcal A}

\newcommand{\Oh}{\mathcal O}
\newcommand{\Ve}{\mathcal V}
\newcommand{\Z}{\mathbb{Z}}

\newcommand{\E}[1]{\left \langle #1 \right \rangle}

\newcommand{\const}{\mathrm{const.}}
\renewcommand{\Re}{\mathrm{Re}}

\begin{document}

\title{State Diagram of the Non-Reciprocal Cahn-Hilliard Model and the Effects of Symmetry}
\author{Martin Kjøllesdal Johnsrud\,\orcidlink{0000-0001-8460-7149}}
\affiliation{Max Planck Institute for Dynamics and Self-Organization (MPI-DS), D-37077 Göttingen, Germany}
\author{Ramin Golestanian\,\orcidlink{0000-0002-3149-4002}}
\email{ramin.golestanian@ds.mpg.de}
\affiliation{Max Planck Institute for Dynamics and Self-Organization (MPI-DS), D-37077 Göttingen, Germany}
\affiliation{Rudolf Peierls Centre for Theoretical Physics, University of Oxford, Oxford OX1 3PU, United Kingdom}
\date{\today}

\begin{abstract}
Interactions between active particles may be non-reciprocal, breaking action-reaction symmetry and leading to novel physics not observed in equilibrium systems. The non-reciprocal Cahn-Hilliard (NRCH) model is a phenomenological model that captures the large-scale effects of non-reciprocity in conserved, phase-separating systems.
In this work, we explore the consequences of different variations of this model corresponding to different symmetries, inspired by the importance of symmetry in equilibrium universality classes.
In particular, we contrast two models, one with a continuous $\mathrm{SO}(2)$ symmetry and one with a discrete $\mathrm{C}_4$ symmetry.
We analyze the corresponding models by constructing three-dimensional linear stability diagrams. 
With this, we connect the models with their equilibrium limits, highlight the role of mean composition, and classify qualitatively different instabilities.
We further demonstrate how non-reciprocity gives rise to out-of-equilibrium steady states with non-zero currents and present representative closed-form solutions that help us understand characteristic features of the models in different parts of the parameter space.
\end{abstract}

\maketitle

\section{Introduction}

\begin{figure*}[t]
\centering
\includegraphics[width=.8\textwidth]{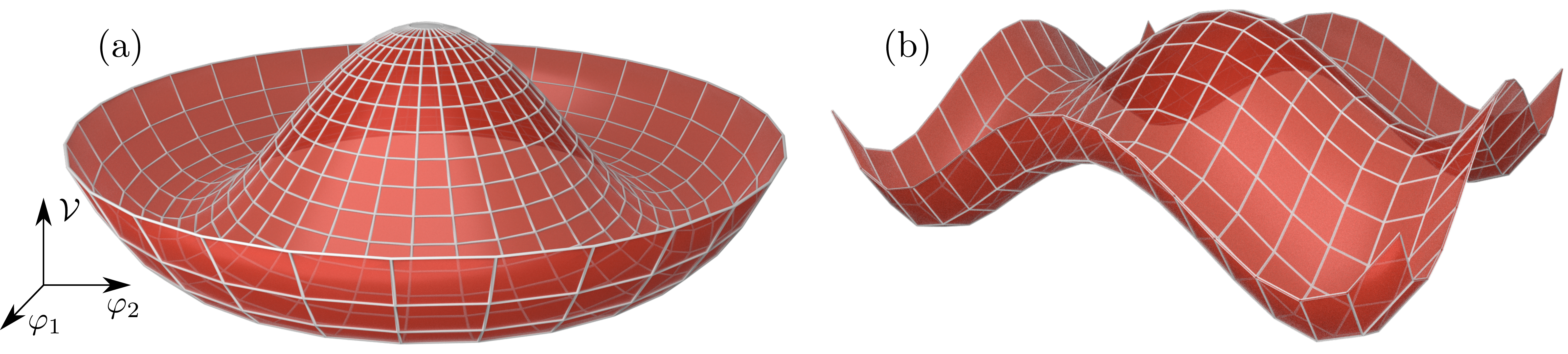}
\caption{(a) The Mexican hat potential of the model with continuous $\mathrm{SO}(2)$ symmetry. (b) The potential of the model with discrete $\mathrm{C}_4$ symmetry.}
	\label{fig: potentials}
\end{figure*}

The physical properties of active matter---comprising soft matter systems, living or artificial, which continually consume energy---have been the subject of extensive investigations in recent decades~\cite{ramaswamyMechanicsStatisticsActive2010,golestanianPhoreticActiveMatter2022,chateDryAligningDilute2022,gompper2020MotileActive2020,cates-nardini-2024}. Depending on the mechanisms that underlie the non-equilibrium activity, some of the very basic properties that apply to passive matter will no longer hold. In particular, the action-reaction symmetry exemplified by Newtons third law relies on forces derived from an underlying common interaction potential, something one cannot expect in effective descriptions of many active systems such as living organisms. Such \emph{non-reciprocal} interactions have recently been a topic of interest in active matter, as it is an important feature of effective chemically-mediated~\cite{sotoSelfAssemblyCatalyticallyActive2014,sotoSelfassemblyActiveColloidal2015,agudo-canalejoActivePhaseSeparation2019,sahaPairingWaltzingScattering2019,meredithPredatorPreyInteractions2020} or hydrodynamically-mediated~\cite{Uchida2010,Hickey2023} interactions, the social interaction of swarms~\cite{fruchartNonreciprocalPhaseTransitions2021}, or in models of the brain~\cite{derridaExactlySolvableAsymmetric1987,sompolinskyTemporalAssociationAsymmetric1986}. Moreover, it has been proposed that naturally occurring non-reciprocal interactions in chemically active systems may have played a role in the emergence of early forms of life \cite{OuazanReboul2023,VincentPRL,OuazanReboul2023NJP}.
A similar action-reaction asymmetry, exhibited by condensed living matter, is \emph{odd} elasticity or viscosity~\cite{scheibnerOddElasticity2020,fruchartOddViscosityOdd2023,tanOddDynamicsLiving2022}.
Non-reciprocal interactions are also present in non-active systems driven out of equilibrium, such as driven dusty plasmas~\cite{melzerStructureStabilityPlasma1996,melzerTransitionAttractiveRepulsive1999,lisinExperimentalStudyNonreciprocal2020,kryuchkovStrangeAttractorsInduced2020}, magnetic systems~\cite{hanaiPhotoinducedNonreciprocalMagnetism2025,nadolnyNonreciprocalSynchronizationActive2025} or Bose-Einstein condensates~\cite{belyanskyPhaseTransitionsNonreciprocal2025}.
More generally, introducing non-reciprocal interactions is one of the simplest way to break equilibrium, making it useful as a theoretical tool for exploring the behavior non-equilibrium states of matter~\cite{aransonWorldComplexGinzburgLandau2002,fruchartNonreciprocalPhaseTransitions2021,Osat2022,youngNonequilibriumUniversalityNonreciprocally2024,johnsrudFluctuationDissipationRelations2025,johnsrudFluctuationDissipationRelations2025a,avniNonreciprocalIsingModel2025,Osat2024}.

Non-reciprocal systems, or more generally non-equilibrium systems, may exhibit \emph{exceptional points}~\cite{katoPerturbationTheoryLinear1995,heissPhysicsExceptionalPoints2012}. These were first considered as features of quantum systems, appearing in models of quantum chaos~\cite{heissTransitionalRegionsFinite1991}, $\mathcal{PT}$-symmetric systems~\cite{benderRealSpectraNonHermitian1998,konotopNonlinearWaves$mathcalPT$symmetric2016}, optical wave-guides~\cite{suchkovNonlinearSwitchingSolitons2016}, and atomic orbitals~\cite{cartariusExceptionalPointsAtomic2007}; all systems where gain and loss result in non-Hermitian terms in the Hamiltonian.
In non-reciprocal systems, exceptional points appear in the linearized dynamics of the system, represented by the linear stability matrix~\cite{fruchartNonreciprocalPhaseTransitions2021,sahaScalarActiveMixtures2020,youNonreciprocityGenericRoute2020}.
This matrix is in general non-Hermitian---i.e., it does not equal its transposed complex-conjugate---due to the broken action-reaction symmetry, and there may therefore be points in parameter space where it becomes defective, its eigenvalues complex, and the corresponding eigenvectors coalesce\footnote{Since the inner product in this case is strictly $L^2$ (including not only sum over components but also integral over the domain), the Hermitian-conjugate also includes integration by parts.}.
Exceptional points in non-equilibrium physics promise to be a fruitful avenue for novel physical effects, such as the recently demonstrated critical fluctuations in driven-dissipative Bose-Einstein condensates~\cite{hanaiCriticalFluctuationsManybody2020} or non-equilibrium extensions of $\text{O}(N)$ models~\cite{zelleUniversalPhenomenologyCritical2024,youngNonequilibriumFixedPoints2020,youngNonequilibriumUniversalityNonreciprocally2024}.

In equilibrium statistical physics, the Cahn-Hilliard model is a widely used phenomenological model for phase separation~\cite{cahnFreeEnergyNonuniform1958,cahnFreeEnergyNonuniform1959}.
The model describes the dynamics of conserved scalar fields $\varphi_a(\bm x, t)$ relaxing to minimize its free energy $F[\varphi]$.
The Non-Reciprocal Cahn-Hilliard (NRCH) model extends this model by introducing non-reciprocal interactions that cannot be described in terms of minimizing free energy~\cite{sahaScalarActiveMixtures2020,youNonreciprocityGenericRoute2020}.
This model has recently been shown to exhibit phase behavior beyond that of its well-studied reciprocal counterpart~\cite{sahaScalarActiveMixtures2020,youNonreciprocityGenericRoute2020,frohoff-hulsmannSuppressionCoarseningEmergence2021,pisegnaEmergentPolarOrder2024,ranaDefectSolutionsNonreciprocal2024}.
As a phenomenological model, NRCH has \textit{a priori} many independent parameters. In the literature, many different versions of NRCH have been studied based on the specific focus of the intended investigation, such as a version where the two species are treated on an equal footing \cite{sahaScalarActiveMixtures2020}, a version where one species has minimal dynamics \cite{youNonreciprocityGenericRoute2020}, and a version where rotational invariance in the space of species is introduced \cite{sahaScalarActiveMixtures2020,pisegnaEmergentPolarOrder2024,ranaDefectSolutionsNonreciprocal2024,sahaEffervescenceBinaryMixture2025}.
In this paper, we aim to investigate the role of symmetry with respect to transformations in the space of fields on the behavior of the NRCH model, and take the first steps towards categorizing the phenomenology of its different variations.

To explore the consequences of symmetry, we consider two versions of the model: one with the discrete $\mathrm{C}_4$-symmetry and one with the continuous $\mathrm{SO}(2)$ symmetry (see \autoref{fig: potentials}).
These restrictions allow us to fully characterize their state diagram at the linear level.
The symmetries are inherited by the linear-stability diagrams, reducing the number of free parameters in the continuous case. In the discrete case, an energy barrier leads to distinct phases separated by a finite thickness domain wall, while in the continuous case there is no barrier, leading to ``continuous phase separation''. We further explore how these effects persists even when the system is driven out of equilibrium by non-reciprocal interactions.

\section{The Model}

The model we consider consists of two real scalar fields, $\varphi_a(\bm x, t)$, $a \in \{1, 2\} $.
These fields represent a density fluctuation at time $t$ and position $\bm x$.
The reciprocal interactions are due to a free energy density,
\begin{equation}
	\Eff[\varphi] = \frac{1}{2}  \bm \nabla \varphi_a \cdot \bm \nabla \varphi_a  + \Ve(\{\varphi_a\}),
\end{equation}
where summation over repeated indices is implied. The total free energy reads $F = \int \dd \bm x \, \Eff$.
The two models are defined by their potential $\Ve(\varphi)$.
The potential of the model with $\mathrm{SO}(2)$ symmetry is
\begin{equation}
    \Ve_\mathrm{cs}(\{\varphi_a\}) = \frac{r}{2} \varphi_a \varphi_a + \frac{u}{4} (\varphi_a \varphi_a)^2.
\end{equation}
If $r \geq 0$, the potential has one global minimum at $\varphi_a = 0$, whereas for $ r< 0$ the potential takes the characteristic Mexican hat shape illustrated in \autoref{fig: potentials}. The potential is a function of $\varphi_a \varphi_a$ only, which is a construction that is invariant under rotations and reflections (e.g. $\varphi_{1} \rightarrow - \varphi_1$) in the field space. These transformations form the Lie group $\mathrm{O}(2)$.
The dynamics arising from this potential corresponds to Model B in the classification of Halperin and Hohenberg~\cite{hohenbergTheoryDynamicCritical1977}.
The symmetry is reduced to $\mathrm{SO}(2)$ with the introduction of a non-reciprocal term $\alpha \epsilon_{ab}\varphi_b$ in the equations of motion (see below).
This reduction cannot happen at the potential level, without also breaking rotational invariance, which is why the potential has a larger symmetry group than the model.

To represent an example with {\em discrete symmetry}, we choose the potential
\begin{align}
    \Ve_{\mathrm{ds}}(\{\varphi_a\}) = \sum_a  \left[\frac{r}{2} \varphi_a^2 + \frac{u}{2}\varphi_a^4\right],
\end{align}
which is not invariant under continuous rotations, as illustrated in \autoref{fig: potentials}. 
However, $\Ve_{\mathrm{ds}}$ is invariant under operations in $\mathrm{D}_4$,
the dihedral group of order 8 corresponding to the symmetries of a square.
These operations are rotations with $n \pi / 2$ radians in addition to reflections, e.g. $\varphi_1 \rightarrow - \varphi_1$. This, similarly to the the symmetry group of the single-field Cahn-Hilliard [$\mathrm{O}(1) \cong \mathrm{D}_2]$, is a discrete group and therefore gives rise to a phase separation with a well-defined domain wall.
The introduction of a non-reciprocal term restricts the symmetry to $\mathrm{C}_4$, namely, the cyclic group of order 4.

\begin{figure*}[t]
\includegraphics[width=.45\textwidth]{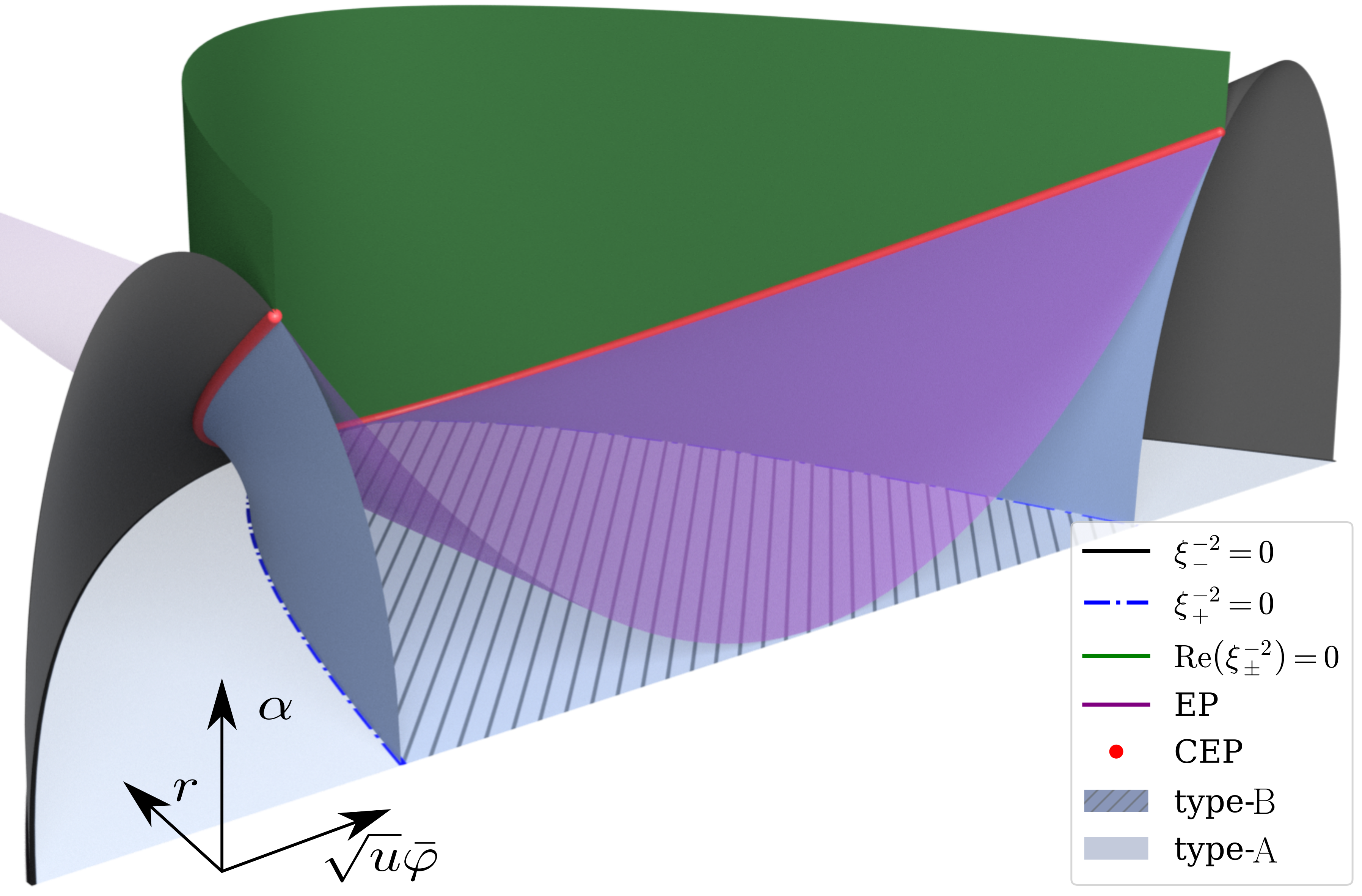}
\includegraphics[width=.54\textwidth]{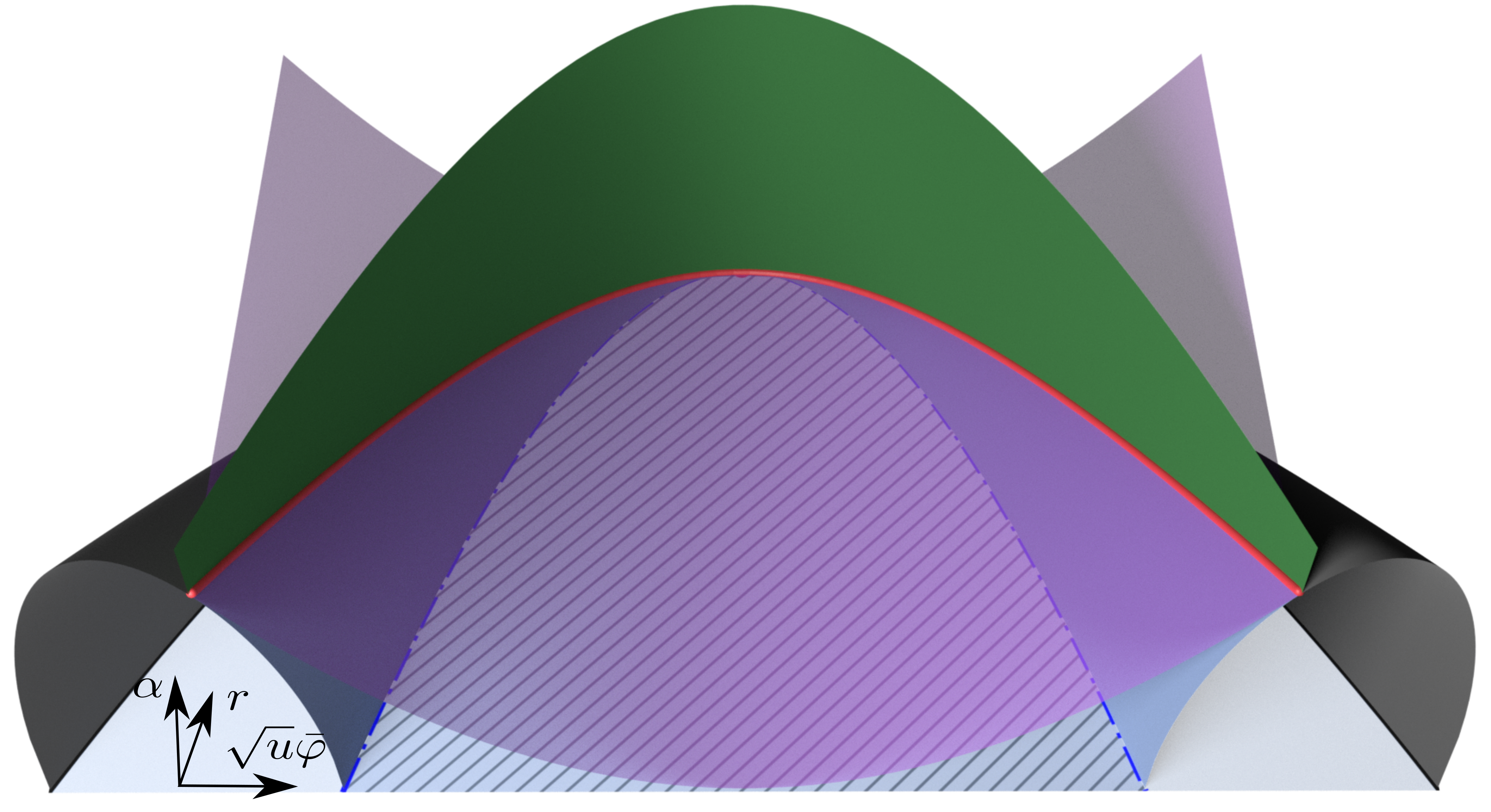}
\caption{
The state diagram of the model with continuous symmetry.
The exceptional plane is highlighted in purple, and the planes between the stable, type-A unstable, and type-B unstable are highlighted in black, blue, and green, respectively. The red line is the intersection of all these surfaces.
}
  \label{fig: phase surface}
\end{figure*}

In an equilibrium model, the chemical potential of field $\varphi_a(\bm x, t)$ is $\mu_a(\bm x, t) = \smash{\fdv{F[\varphi]}{{\varphi_a(\bm x, t)}}}$. This definition implies $M_{ab}\delta(\bm x - \bm x') = \smash{\fdv{\mu_a(\bm x, t)}{{\varphi_b(\bm x', t)}}}$ is symmetric. This matrix represents the thermodynamic forces acting {\em on} field (species) $a$ {\em by} field (species) $b$, and the symmetry of the matrix implies that the interactions are reciprocal. We now add a term $\tilde \mu_a$ to the chemical potential that gives rise to an anti-symmetric part of $M_{ab}$, thus corresponding to non-reciprocal interactions. By stipulation, such a term cannot be derived from a free energy. 
In both models, we introduce a minimal non-reciprocal term $\tilde \mu_a = \alpha \epsilon_{ab}\varphi_b$ so that $\smash{\fdv{{\tilde\mu_a(\bm x, t) }}{{\varphi_b(\bm x', t)}}}= \alpha \epsilon_{ab}\delta(\bm x-\bm x')$, where $\epsilon_{ab}$ is the anti-symmetric Levi-Civita tensor, with $\epsilon_{12} = -\epsilon_{21} = 1$.
This coupling has recently been studied in~\cite{sahaScalarActiveMixtures2020,youNonreciprocityGenericRoute2020}, and more general non-reciprocal extensions have been proposed in Ref.~\cite{sahaEffervescenceBinaryMixture2025}.
The field $\varphi_a$ obeys the conservation law 
\begin{equation}
\partial_t \varphi_a + \bm\nabla \cdot \bm j_a=0,    
\end{equation}
where the flux $\bm j_a$ is given by thermodynamic forces (affinities) that arise from gradients in the chemical potential, namely, ${\bm j}_a = -\Gamma {\bm \nabla}\mu_a$, with $\Gamma$ being a mobility.
The governing dynamical equations thus read
\begin{equation}
  \label{eom}
	\partial_t \varphi_a 
	= \Gamma \nabla^2 
    \left[ 
        \pdv{\Ve}{\varphi_a} -\nabla^2 \varphi_a  + \alpha \epsilon_{ab}\varphi_b 
    \right].
\end{equation}

\section{Linear Stability}

We now construct linear stability diagrams for the two cases that represent different symmetries by using linear stability analysis.
We study the response of the system to small perturbations around a homogeneous state, represented by $\bar \varphi_a$.
Inserting $\varphi_a(\bm x, t) = \bar \varphi_a + \delta \varphi_a(\bm x, t)$ into the equations of motion, we obtain the leading order contribution in Fourier space as follows
\begin{equation}
  \partial_t \delta \varphi_a(\bm q, t)
  = -  q^2 \Gamma M_{ab}(\bm q) \delta \varphi_{b}(\bm q,t).
\end{equation}
The matrix $M_{ab} $ depends on the specific choice for the potential, and thus reflects the effects of the underlying symmetry.
We set $\Gamma = 1$, which can be done without loss of generality by scaling time $t$.

A non-zero value of $\bar \varphi_a$ seemingly breaks the assumed symmetry of the model.
However, the symmetry is still present in the form of rotation of both $\bar \varphi$ and $\delta \varphi$.
In the case of the continuous symmetry the Ward-Takahashi identities related to rotation still apply~\cite{johnsrudRenromalizationTBP}, just as they do when this symmetry is broken spontaneously.
It might appear tempting to invoke Goldstone's theorem in this case, treating $\bar \varphi$ as the order-parameter whose non-zero value may indicate a broken symmetry giving rise to slow modes.
However, as the system already features two slow-modes due to the conservation laws, the breaking of the continuous symmetry of $\varphi$-rotation does not yield any additional slow-modes, and thus Goldstone's theorem does not provide any information about the dynamics of the system\footnote{In terms of the vertex-functions $\Gamma_{ab}(\omega, \bm q)$---the inverse Greens-functions---Goldstone's theorem states that $\E{\varphi_a} \Gamma_{ab}(\bm q= \omega=0) = 0$~\cite{tauberCriticalDynamicsField2014}.
So, either the symmetry is preserved ($\E{\varphi_a} = \bar \varphi_a = 0$), or $\Gamma$ has a zero.
However, the conservation laws ensure that $\Gamma(\omega = 0) \propto q^2$, independently of whether or not the symmetry is broken.}.
In fact, the value of $\bar \varphi_a$ is a fixed parameter, and not chosen spontaneously by the system.

\subsection{Continuous Symmetry}

\begin{figure*}[t]
\includegraphics[width=.85\textwidth]{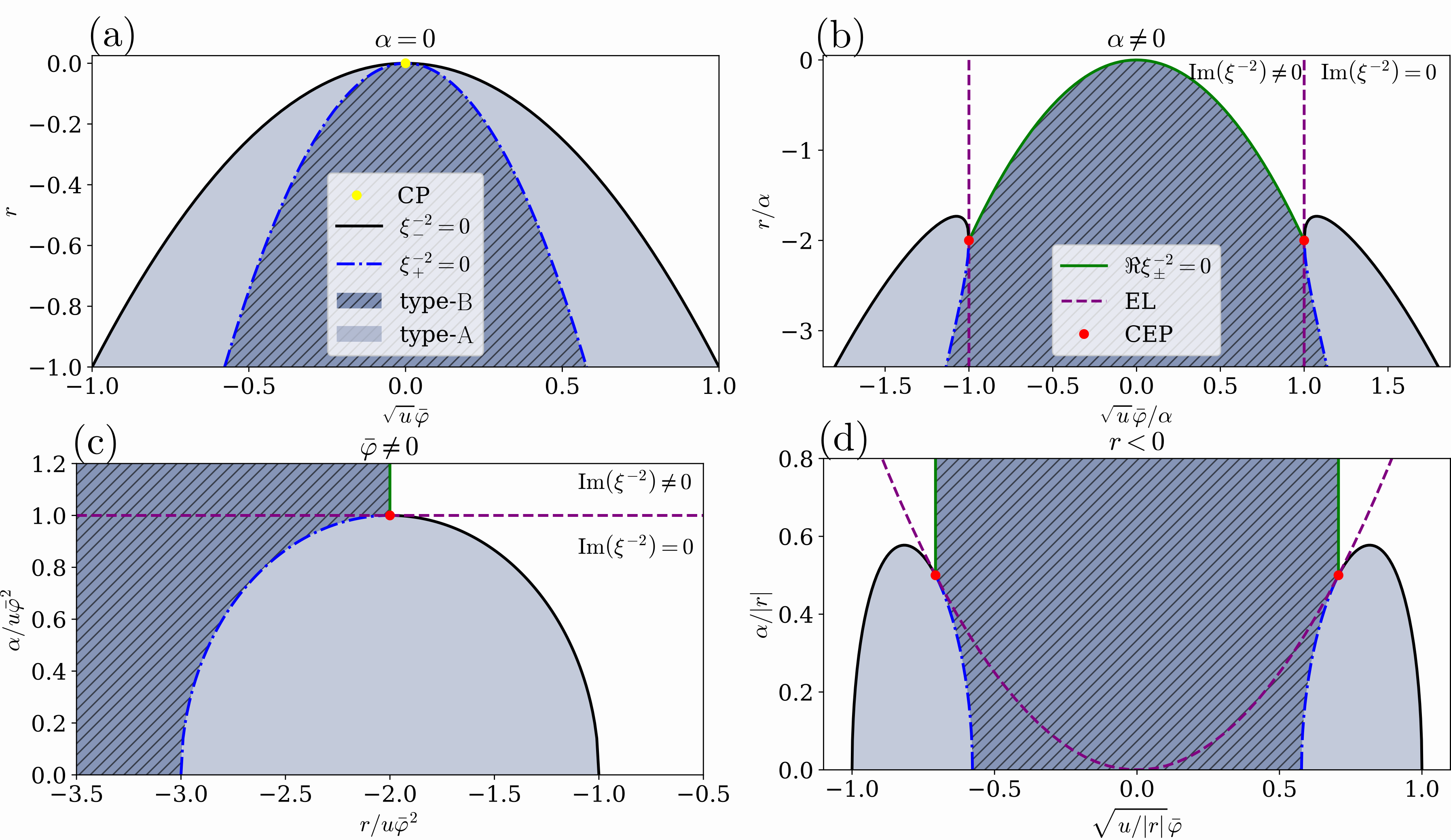}
\caption{
The four non-trivial slices of the full three-dimensional state diagram shown in \autoref{fig: phase surface}.
Panel (a) corresponds to the equilibrium case $\alpha = 0$, while panels (b)-(d) illustrate three orthogonal slices for finite $\alpha$, at constant $\alpha$, $\bar \varphi$, and $r$, respectively. Light blue and hashed areas mark the type-A and type-B unstable regions. 
Black, blue dash-dotted, and green lines separate stable and type-A, type-A and type-B, and stable and type-B, respectively.
The purple dashed line marks the exceptional points, the yellow dot the critical point, and the red dots the critical exceptional points where all the lines intersect.
}
	\label{fig: diagrams}
\end{figure*}

We first analyze the rotationally invariant potential, choosing $\bar\varphi_a = (\bar\varphi, 0)$ without loss of generality, which yields
\begin{equation}
  M_{ab}
  =
  \begin{pmatrix}
      q^2 + r + 3 u \bar \varphi^2&
      \alpha \\
      - \alpha&
      q^2 + r + u \bar \varphi^2
  \end{pmatrix}.
\end{equation}
The corresponding eigenvalues, namely,
\begin{equation}
  \lambda_\pm = q^2 + r + 2u\bar\varphi^2 \pm \sqrt{ ( u \bar\varphi^2)^2 - \alpha^2 },
\end{equation}
encode the stability of the homogeneous solution. A perturbation $\delta \varphi$ will decay if the real part of the corresponding eigenvalue is positive, implying that the homogeneous state $\bar \varphi$ is stable.
Since the $q^2$ term always contributes positively, the least stable eigenvalues can be obtained via $\xi^{-2}_\pm \equiv \lambda_\pm(q \rightarrow 0)$, which can be used to define correlation lengths $\xi_\pm$ in terms of the three free parameters, $r$, $\sqrt{u}\bar \varphi$, and $\alpha$.
For $r > 0$, all values of $\bar\varphi$ result in stable solutions. Therefore, to look for instabilities, we set $r\leq0$.
The critical region defined by $\Re\left(\xi_\pm^{-2}\right) = 0$, therefore, forms a two-dimensional surface in the three-dimensional parameter space. This is illustrated by the surfaces in \autoref{fig: phase surface}. Additional angles are shown in the Supplementary Material \cite{SupplementalMaterial}.

The base of this diagram is the $\alpha = 0$ surface. We choose $\alpha > 0$, without loss of generality.
At $\alpha = u \bar \varphi^2$, we have $\xi^{-2}_+ = \xi^{-2}_-$.
In reciprocal systems, this indicates a degeneracy where two independent eigenmodes have the same eigenvalue.
When we deal with non-reciprocal systems, and thus non-Hermitian matrices, we are no longer guaranteed that the eigenvectors of $M_{ab}$ span the full vector space of perturbations. At this exceptional point (EP) \cite{katoPerturbationTheoryLinear1995,heissExceptionalPointsTheir2004}, the eigenvalues coincide because the two eigenmodes align, rendering the matrix $M_{ab}$ defective. As there are three free parameters and the exceptional points are given by one equation, namely, $\alpha = u\bar \varphi^2$, the full set of exceptional points forms a 2D surface, which we denote as the exceptional plane. This is illustrated in purple in the \autoref{fig: phase surface}.
Above the exceptional plane, the two eigenvalues become complex, and the perturbations will no longer decay or diverge monotonically: they exhibit oscillatory behavior. Here, we have $\Re\left(\xi_\pm^{-2}\right) = r + 2 u \bar \varphi^2$, and, therefore, both modes transition from stable to unstable at the same surface, defined via $r = - 2 u \bar \varphi^2$.
The intersection between the critical and the exceptional planes can be called critical exceptional points (CEPs) \cite{hanaiCriticalFluctuationsManybody2020}.
In the dynamical systems literature, these are known as Takens-Bogdanov bifurcations~\cite{guckenheimerNonlinearOscillationsDynamical1983}.
These CEPs are parameterized as follows
\begin{equation}
  \left(\sqrt{u} \bar\varphi, r, \alpha\right) 
  = \left(s, -2 s^2, s^2\right), \quad s \in \mathbb{R},
\end{equation}
and illustrated by the red line in \autoref{fig: phase surface}. On this line, the two eigenmodes align and are both critical.

To better present the linear stability diagram, we now examine various two-dimensional slices that can be obtained by fixing one of the parameters. Setting $\alpha = 0$ gives the diagram of Model B for $n=2$ fields, with $\xi_\pm^{-2} = r + (2\pm1) u\bar\varphi^2$.
This is illustrated in \autoref{fig: diagrams}(a), and is the base of the diagram in \autoref{fig: phase surface}.
At $r = \bar\varphi = 0$, the correlation lengths diverge.
This is a critical point corresponding to the onset of phase separation.
Starting in the white region of the diagram, where $u \bar\varphi^2> |r|$, we have $\xi_\pm^{-2} > 0$, and thus both modes are stable.
Upon further lowering $\bar\varphi$, we reach a line in the diagram at $u \bar \varphi^2 = |r|$---shown as a black line in \autoref{fig: phase surface}---where $\xi^{-2}_- = 0$ and the corresponding eigenmode transitions from stable to unstable. Therefore, perturbations in this direction in $\varphi$-space ($\delta \varphi_2$ in this case) will grow exponentially.

This diagram is similar to the binodal-spinodal diagram for phase separation of one species.
In that case, the outer line represents the binodal, within which the  the system is metastable, while the inner line gives the spinodal, where the system becomes linearly unstable.
In this case, on the other hand, the system has an additional soft direction (orthogonal to the radial direction) as a result of the rotational symmetry. The outer line (the type-A border) therefore represents the onset of linear \emph{instability}, while also being the binodal, and no metastable state exists.
The inner line (the type-B border) is where the second eigenvalue changes sign.
At $r = \bar \varphi = \alpha = 0$ is the well-known $\mathrm{O}(2)$ critical point.
For $\alpha\neq0$, the outer line denotes linear instabilities, but the identifying binodals becomes more complicated~\cite{greveCoexistenceUniformOscillatory2025,sahaPhaseCoexistenceNonreciprocal2024}.
For small $\alpha$, we expect the instabilities to be a good approximation for the binodals.

Looking back at the Mexican hat potential illustrated in \autoref{fig: potentials}, this corresponds to a point inside the ring of minima, and perturbations that would correspond to the Goldstone mode in a spontaneous symmetry breaking scenario (orthogonal to the radial direction). The mode corresponding to perturbations in the radial direction, $\delta \varphi_1$, remains stable. This holds in the light blue area labeled ``type-A unstable'', where there is one unstable mode. At $u \bar \varphi^2 =|r|/3$, illustrated by a blue dash-dotted line, the second mode also transitions from stable to unstable. The area where both modes are unstable is highlighted as dark blue and hashed, with the label ``type-B unstable''.

Assuming all parameters are non-zero, we may re-scale the correlation lengths by either of them, thus eliminating one parameter. For example, $\xi^{-2}_\pm/\alpha$ depend only on $r/\alpha$ and $\sqrt{u}\bar \varphi/\alpha$. Any two slices where $\alpha = \const \neq 0$ are thus equivalent as they are related by a scale transformation.
There are therefore three additional
non-trivial two-dimensional slices, illustrated by  in \autoref{fig: diagrams}(b)-(d).
Their intersection with the three-dimensional surface is shown in Fig. S3 in the SM~\cite{SupplementalMaterial}.

We first focus on the $\alpha = \const$ slice, illustrated in \autoref{fig: diagrams}(b).
The equations for the critical lines of the real eigenvalues are
\begin{align}
  \frac{r}{\alpha} &= - 2 \,\frac{u\bar\varphi^2}{\alpha} \pm \sqrt{\frac{u^2\bar\varphi^4}{\alpha^2} - 1}, &
   u |\bar\varphi|^2 &> \alpha.
\end{align}
New to this diagram are the exceptional lines $\sqrt{u} \bar \varphi = \pm \alpha$, illustrated by purple dashed lines in \autoref{fig: diagrams}(b).
Outside these lines, the diagram is similar to the $\alpha = 0$ diagram, while between them, the correlation lengths become complex and the region with one stable mode vanishes.
The critical line is here given by
\begin{align}
  r &= - 2 u \bar \varphi^2, & u |\bar\varphi|^2 &\leq \alpha,
\end{align}
and is highlighted in green, in \autoref{fig: diagrams}(b), while the intersections of this line with the critical exceptional lines are illustrated as red dots.

Next, we consider the $\sqrt{u}\bar \varphi = \mathrm{const.}$ slice, illustrated in \autoref{fig: diagrams}(c).
Below the exceptional line, the critical lines form two quarter-circles, which meet at the critical exceptional point, defined by
\begin{equation}
  \left( \frac{r}{u\bar\varphi^2} + 2 \right)^2 + \left(\frac{\alpha}{u\bar\varphi^2}\right)^2 = 1,
\end{equation}
while above the exceptional line, the critical line is given by the vertical line $r/(u\bar\varphi^2) = - 2 $.

Finally, the slice where $r=\mathrm{const.}$ (as normalized by $|r|$) is illustrated in \autoref{fig: diagrams}(d). The equation for the critical lines below the exceptional line is
\begin{equation}
  \frac{\alpha}{|r|} = \sqrt{-1 + 4\,\frac{u\bar\varphi^2}{|r|} - 3\left(\frac{u\bar\varphi^2}{|r|}\right)^2},
\end{equation}
while above it, the critical line is again given by vertical lines, defined by $\sqrt{u/|r|} \bar \varphi = \pm 1/ \sqrt{2}$.

\subsection{Discrete Symmetry}

\begin{figure*}[t]
  \centering
  \includegraphics[width=.85\textwidth]{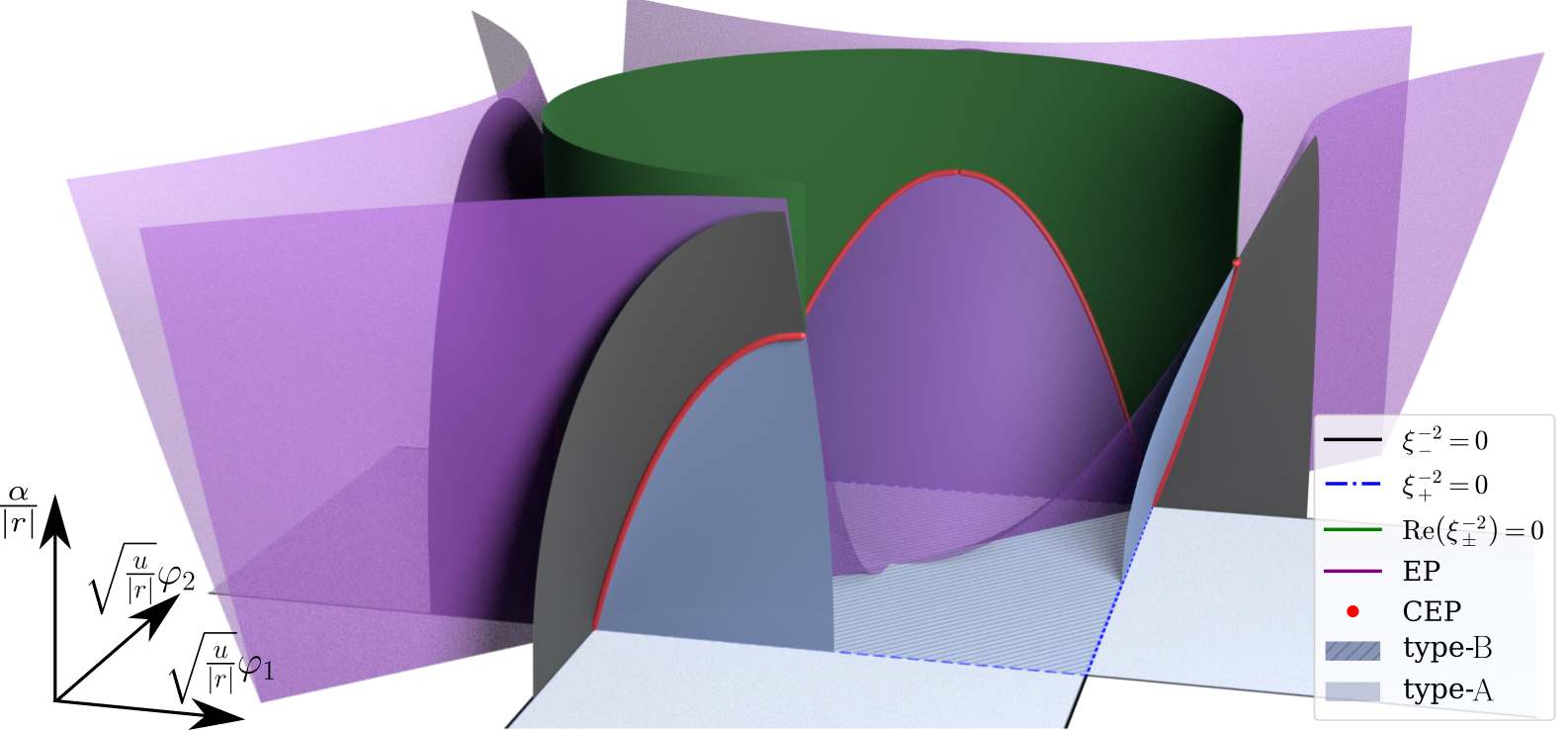}
  \caption{
    The state diagram of the model with discrete
    symmetry, scaled by $|r|$.
    Light blue and hashed areas mark the type-A and type-B unstable regions. 
    Black, blue, and green planes separate stable and type-A, type-A and type-B, and stable and type-B regions, respectively.
    The exceptional plane is highlighted in purple, and the red line marks the critical exceptional points where all the planes intersect.
  }
  \label{fig: phase 3d assym}
\end{figure*}

\begin{figure*}[t]
  \centering
  \includegraphics[width=\textwidth]{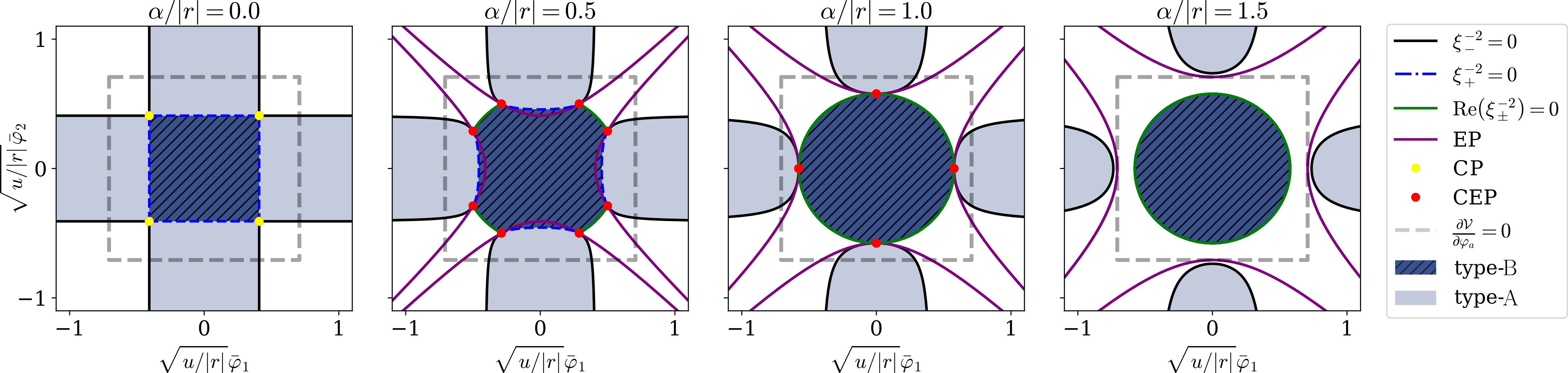}
  \caption{
    Constant $\alpha$ slices of the linear stability diagram illustrated in \autoref{fig: phase 3d assym}. 
    The gray dashed lines represent the line connecting the minima of the potential illustrated in \autoref{fig: potentials}(a), and form a square centered at the origin and with side lengths $\sqrt{u/|r|} \bar \varphi = \sqrt{2}$.
    Light blue and hashed areas mark the type-A and type-B unstable regions. 
    Black, blue dash-dotted, and green lines separate stable and type-A, type-A and type-B, and stable and type-B, respectively.
    The purple lines mark the exceptional points, the yellow dots the critical point, and the reds dot the critical exceptional points where all the lines intersect.
  }
  \label{fig: assym}
\end{figure*}

We now apply this analysis to the model with a square potential. The linear stability matrix is then
\begin{equation}
  M_{ab}
  = 
  \begin{pmatrix}
      q^2 + r + 6 u \bar \varphi_1^2 & \alpha \\
      - \alpha & q^2 + r + 6 u \bar \varphi_2^2
  \end{pmatrix},
\end{equation}
which has the eigenvalues 
\begin{equation}
  \lambda_{\pm} = q^2 + r + 3 u \left(\bar \varphi_1^2 + \bar \varphi_2^2\right) \pm \sqrt{9 u^2 (\bar \varphi_1^2 - \bar \varphi_2^2)^2 - \alpha^2}.
\end{equation}
Since the eigenvalues depend on both $\bar \varphi_1^2$ and $\bar \varphi_2^2$, the full linear stability diagram requires another dimension to visualize.
For this model,
we focus on a 3D slice of the full 4D diagram as shown in \autoref{fig: phase 3d assym}. This cross section is related to any other 3D slice via a choice of scaling, which in this case is chosen to be $|r|$.
2D cross sections of \autoref{fig: phase 3d assym} for $\alpha=\const$ is shown in \autoref{fig: assym}.
For a comparison with the continuous case, \autoref{fig: phase 3d assym} corresponds to \autoref{fig: diagrams}(d).
The inverse-squared correlation lengths defined via  $\xi_\pm^{-2} = \lambda_{\pm }(q\rightarrow 0)$ as above are then given as follows
\begin{eqnarray}
  \xi_\pm^{-2}
  &= -|r| + 3 u \left(\bar \varphi_1^2 +\bar \varphi_2^2\right) \nonumber 
  \pm \sqrt{9 u^2 \left({\bar \varphi_1^2} - {\bar \varphi_2^2}\right)^2 - \alpha^2}.
\end{eqnarray}
Where the eigenvalues are real (i.e. outside the exceptional plane) we can find the critical surface by solving $\xi_\pm^{-2} = 0$, e.g. for $\bar\varphi_2$, which yields 
\begin{equation}
  \bar\varphi_2
  =
  \sqrt{\frac{|r|^2 + \alpha^2 - 6 u |r| \bar \varphi_1^2}{6u \left(|r| - 6 u\bar \varphi_1^2\right)}
  },
\end{equation}
with a similar solution for $\bar \varphi_1$. 
See the Supplementary Material~\cite{SupplementalMaterial} for plots of this for a range of parameter values.
The critical surface is highlighted in green in \autoref{fig: phase 3d assym}.

For $\alpha = 0$ the critical lines are straight, where $\sqrt{u/|r|} \bar \varphi_1$ or $\sqrt{u/|r|} \bar \varphi_2$ are $\pm 1/\sqrt{6} \approx \pm 0.4$.
As $\alpha$ is increased, the critical lines become curved, asymptotically approaching the $\alpha = 0$ lines for large values of the fields.
The exceptional plane, shown in purple in \autoref{fig: phase 3d assym}, is given by a set of hyperbolae defined via $u |\bar \varphi_1^2 - \bar \varphi_2^2| = \alpha/3$, where we again have chosen $\alpha > 0$ for concreteness.
Inside the area bounded by the exceptional surface, i.e., the subset containing the origin, the critical surface is given by the circle $u (\bar \varphi_1^2 + \bar \varphi_2^2)=|r|/3$ where $\Re (\xi_\pm^{-2})= 0$, which is shown in green in \autoref{fig: phase 3d assym}.
The intersection of the critical surface and the exceptional surface is the critical exceptional line, which is shown in red in \autoref{fig: phase 3d assym}.
These are located at the intersections of the circle of radius $1/\sqrt{3}$ and the hyperbolas with semi-major axis $\smash{\sqrt{\alpha/(3|r|)}}$. For $\alpha > |r|$ there are no such points, for $\alpha = 0$ and $\alpha = |r|$ there are four, while for $\alpha/|r| \in (0, 1)$ there are eight, corresponding to the elements of $\mathrm{D_4}$. These observations can be seen in the two-dimensional slices of the linear stability diagram shown in \autoref{fig: assym}.
Additional variations of such plots are presented in the supplemental material \cite{SupplementalMaterial}.

\section{Phase separation and NESS}

When a mode is unstable, a perturbation of the homogeneous state will grow exponentially, meaning that the system cannot maintain a homogeneous state, and instead undergoes phase separation.
The $\alpha = 0$ linear stability diagram in \autoref{fig: diagrams} resembles the one-field ($n = 1$) double-well Cahn-Hilliard model~\cite{cahnFreeEnergyNonuniform1958}.
In that case, the hashed inner region is unstable, and is separated by the spinodal to the metastable region $u \bar \varphi^2 \in ( |r|, |r|/3 )$ shaded blue.
Here, perturbations will not grow exponentially, but a phase-separated system will have lower free energy and is thus thermodynamically favored.
In the case of $n = 2$, however, this region is unstable, not metastable.
The absence of a metastable homogeneous state for the $\mathrm{SO}(2)$-model is a consequence of the set of minima of the potential being \emph{connected} due to the spontaneously broken continuous symmetry of the system, in contrast to the $\mathrm{C}_4$ model, which is phenomenology closer to that of the $n = 1$ Cahn-Hilliard.
After adding the non-reciprocal term $\alpha$, the corresponding steady-state may no longer be an equilibrium-like, current-free state, but rather a non-equilibrium steady-state (NESS).
Still, we may apply similar analysis for insight into these states.
We now discuss the phenomenology of phase separation, in and out of equilibrium, for the two cases.

\subsection{Discrete Symmetry}

For $\alpha = 0$, the $\mathrm{C}_4$ system forms phases separated by a domain wall whose thickness is independent of the size of the system.
Take the special case of $\bar \varphi_i = \varphi^*(0, -1)$, where $\varphi^* = \sqrt{|r| / (2 u)}$ is the absolute value of the fields in the minima of the potential, i.e., the average is midway between two minima in the square potential landscape illustrated in \autoref{fig: potentials}.
Due to the decoupled non-linearities of the square model, this system admits an exact solution, in the thermodynamic limit, similar to the well-known domain-wall solution of the one-field Cahn-Hilliard model~\cite{cahnFreeEnergyNonuniform1958}, namely
\begin{subequations}
  \label{domain wall}
  \begin{align}
    \varphi_1(\bm x, t) & = \varphi^* \tanh[(x - x_0)/ \ell], \\
    \varphi_2(\bm x, t) & = - \varphi^*,
  \end{align}
\end{subequations}
where $\ell  = \sqrt{2 / |r|}$ is the domain wall thickness, which results from a trade-off between the surface tension term $\nabla^4 \varphi$ and the potential term $r \nabla^2 \varphi$.
This domain wall profile is, in fact, valid for any average composition $\bar \varphi$.
Since there are four degenerate minima in the potential landscape, as illustrated in Fig.~\ref{fig: potentials}, there may be more than two phases, but the walls between them will still have a hyperbolic tangent profile with the same length scale $\ell$.
See Ref.~\cite{frohoff-hulsmannSuppressionCoarseningEmergence2021} for a more complete discussion.

\begin{figure}[t]
  \centering
  \includegraphics[width=.6\columnwidth]{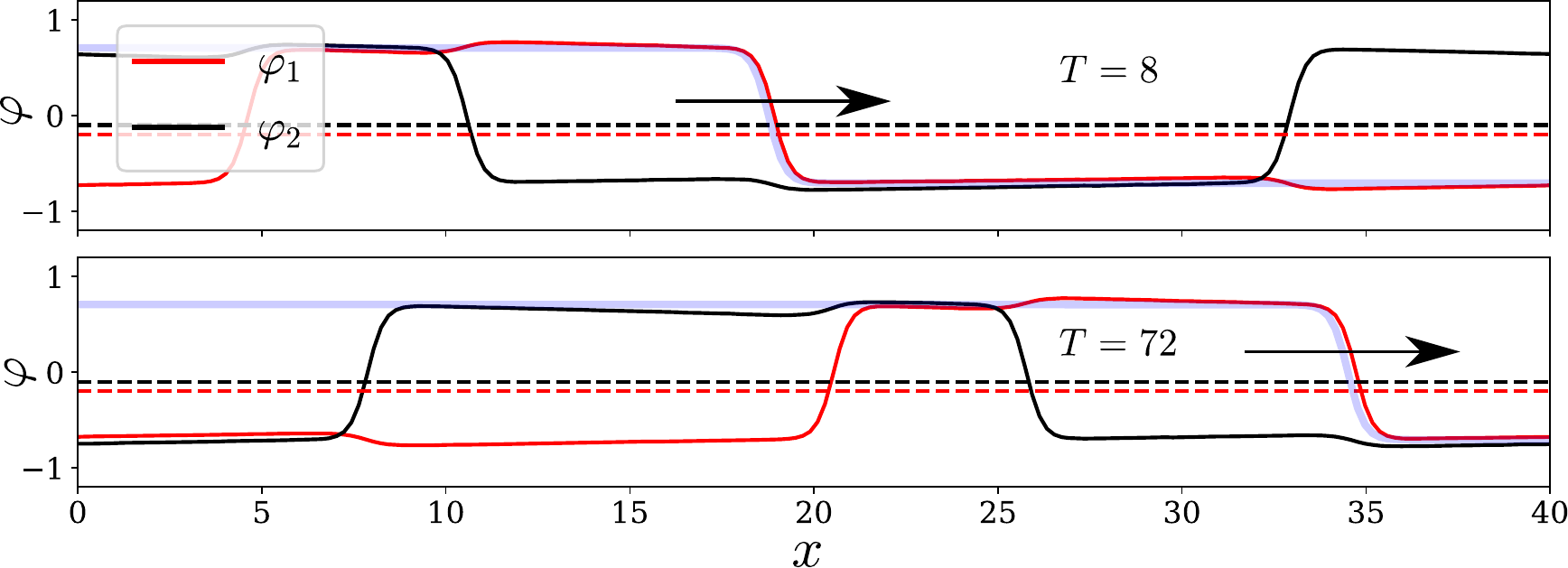}
  \caption{
  Numerical simulation of traveling phase-separated domains in the model with discrete symmetry, for small $\alpha$ and $\bar \varphi$, at two different times $T$.
  The concentrations of the two different species are illustrated in full red and black lines, the dashed lines illustrate the corresponding value for $\bar \varphi$.
  The domain wall of thickness $\ell$, with a hyperbolic tangent profile, is illustrated in light blue.
  The parameters are $u = -r = 10$, $\alpha = 2$, $(\bar \varphi_1, \bar \varphi_2) = -(0.1, 0.2)$.
  }
  \label{fig: Traveling wall}
\end{figure}

For $\alpha \neq 0$, the considerations become more complicated.
As demonstrated recently, the central part of the phase diagram gives rise to traveling wave states~\cite{sahaScalarActiveMixtures2020}.
However, at least where $\alpha$ is not too large, the logic of a trade-off between potential and gradient energy---which gives rise to the finite domain wall---still applies.
As illustrated in \autoref{fig: Traveling wall}, numerical simulations show that states with small $\alpha$ and $\bar \varphi$ still form separate phases, whose traveling domain walls are well approximated by a hyperbolic tangent with a thickness $\ell$.
See \texttt{video\_1.mp4} in~\cite{SupplementalMaterial} for further illustration.

For some restricted versions of the NRCH, one may construct a ``spurious gradient flow''~\cite{sahaPhaseCoexistenceNonreciprocal2024,greveCoexistenceUniformOscillatory2025}.
They balance ``spurious'' pressure and chemical potential across a stationary domain wall; however, this is not available for traveling waves or domain walls.
A model similar to those considered in ~\cite{sahaPhaseCoexistenceNonreciprocal2024,greveCoexistenceUniformOscillatory2025} is obtained by adding a term $- \rho \varphi_1 \varphi_2$ to the $\mathrm{C}_4$ model, breaking its symmetry down to $\mathbb{Z}_2$.
As seen in recent investigations~\cite{braunsNonreciprocalPatternFormation2024}, in models with only $\mathbb{Z}_2$-symmetry the onset of traveling-wave states happens for finite non-reciprocity~\cite{sahaScalarActiveMixtures2020,suchanekEntropyProductionNonreciprocal2023,braunsNonreciprocalPatternFormation2024}---when the non-reciprocal coupling $\alpha$ is larger than the reciprocal coupling $\rho$ and the system passes an exceptional point.
This thus illustrates how symmetries allow us to achieve steady-states with a non-zero current without the need to turn up the non-reciprocal coupling beyond a finite threshold.
We can relate this behavior to the symmetry of the model\footnote{
Here, the special point corresponds to the reciprocal coupling $\rho$ vanishing.
This corresponds to a symmetry $\Z_2\times\Z_2$ under $\varphi_a \rightarrow - \varphi_a$, $\alpha \rightarrow - \alpha$.}.

To illustrate the quantitative implications of this point, let us consider the entropy production. In the formalism of non-equilibrium thermodynamics~\cite{grootNonequilibriumThermodynamics1984,pottierNonequilibriumStatisticalPhysics2010}, the density of entropy-production $\dot \sigma$ is given by a product of the currents, which appear in the conservation laws $\partial_t \varphi_a = - \bm \nabla \cdot \bm j_a$, and the affinities that drive the currents, namely, $\bm \A_a = - \Gamma D^{-1} \bm \nabla \mu_a$. The total entropy production is found as
$
  \dot {\mathcal{S}} = \int \dd \bm x \, \dot \sigma =  \int \dd \bm x\, \bm j_a \cdot \bm \A_a
$.
In our case, the chemical potential has two contributions, $\mu_a = \smash{\fdv{F}{\varphi_a}} + \tilde \mu_a$, so the entropy production rate is
\smash{$
	D \mathcal{\dot S} = - \dot F - \int \dd \bm x \, \partial_t \varphi_a \tilde \mu_a
$.}
In the steady-state, we have $\langle \dot F\rangle = 0$, as there is no entropy production without a non-variational contribution~\cite{nardiniEntropyProductionField2017}. Observing that $\tilde \mu_a = \alpha \epsilon_{ab} \varphi_b$, and that $\partial_t \varphi$ has a term $\propto \alpha$, we infer that $\dot \sigma \propto \alpha^2$, even in a state with no large-scale motion \cite{alstonIrreversibilityNonreciprocalSymmetryBreaking2023}. We may calculate this explicitly in a special case for the continuous model, as we show below. We can thus conclude that the cost of entropy production for achieving large-scale motion is lowered by models with symmetries.

This observation opens up further avenues for investigating the effects of symmetry on the NRCH model. The two models could be linked by adding a term $g \varphi_1^2 \varphi_2^2$, to investigate the transition between the different regimes, and the effects of small symmetry breaking. Furthermore, the $C_4$ model can be extended to $C_n$, and one would expect the to models to converge as $n\rightarrow \infty$.
Another generalization would consider symmetry in models with non-reciprocal interactions between larger number of fields, such as in the models studied in Ref.~\cite{parkavousiEnhancedStabilityChaotic2025}.

\subsection{Continuous Symmetry}
\label{section: solution}

In the case of the Mexican hat potential illustrated in \autoref{fig: potentials}, all points corresponding to $\bar \varphi = \varphi' \equiv \sqrt{|r| / u}$ define the \emph{connected} ground state manifold.
Any two points on this manifold obey a generalization of the common tangent space construction~\cite{brayTheoryPhaseorderingKinetics1994}, and, consequently, the minimum of the free energy will therefore no longer be two homogeneous phases separated by a thin domain wall.
Instead, the field can interpolate between two minimum values taking only values in or near the bottom of the potential. The assumption of two phases with a domain wall separating them is therefore not valid anymore; the system will consist of one continuously changing phase.

\begin{figure}[t]
	\centering
	\includegraphics[width=\columnwidth]{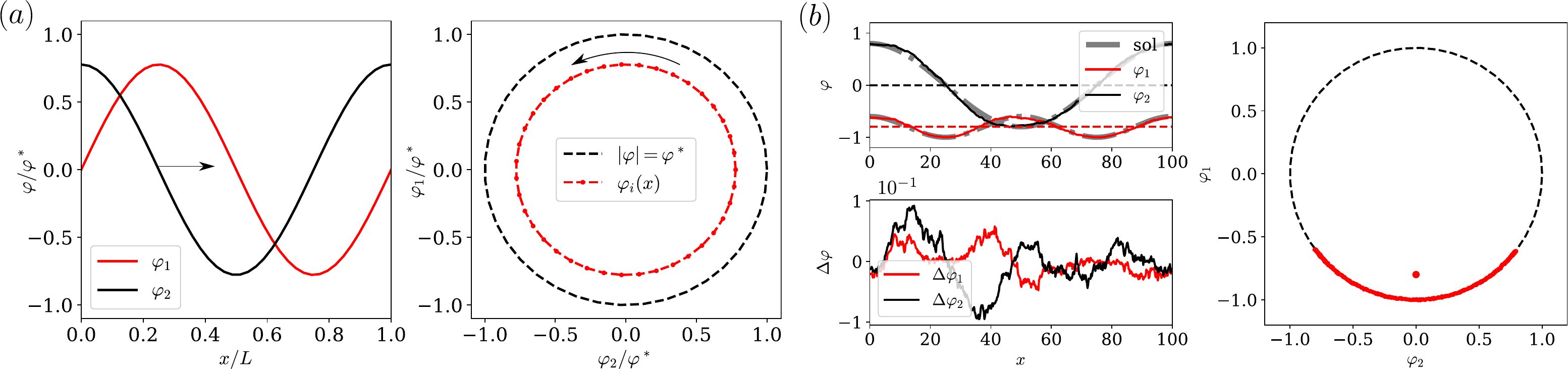}
	\caption{
    (a)
    The traveling wave solution for $\bar \varphi = 0$.
    To the left, the fields are plotted in one dimension of real space, parallel to the wave vector.
    On the right, the same configuration is shown in field space.
    (b)
    The approximate solution for $\bar \varphi \approx \varphi'$ is shown top left in gray, while the numerical solution is overlayed in red and black.
    Bottom left shows the deviation between analytical and numerical solutions, including noise $D = 2\times10^{-4}$.
    The right panel shows the solution in $\varphi_i$ space.
    Parameters are $u = -r = 40$, $(\bar \varphi_1, \bar \varphi_2) = -(0.8,0)$.
    }
	\label{fig: solution 1}
\end{figure}

We can find an exact solution for the $\mathrm{SO}(2)$ model in the case of $\bar \varphi = 0$, corresponding to a slice of the linear stability diagram of the last section where the entire plane, for arbitrary $r<0$, is unstable~\cite{sahaEffervescenceBinaryMixture2025,pisegnaEmergentPolarOrder2024}.
We obtain this solution by making a plane-wave \textit{ansatz} as follows:
\begin{subequations}
\label{solution 1}
\begin{align}
	\varphi_1(\bm x, t; \bm k)
	&= 
	A(k) \sin\phi(\bm x, t; \bm k), \\
	\varphi_2(\bm x, t; \bm k)
	&= 
	A(k) \cos\phi(\bm x, t; \bm k), \\
	\phi(\bm x, t; \bm k) &= \bm k \cdot \bm x - \omega(k) t.
\end{align}
\end{subequations}
Inserting this into the equation of motion Eq.~\eqref{eom} with the rotationally invariant potential, we obtain the following requirements for the solution to be valid
\begin{align}
  \omega(k) &= \alpha k^2 , &
  uA(k)^2 &= - r - k^2.
\end{align}
The function $\phi$ represents the phase of the wave, which depend on $\bm x$ and $t$.
The solution corresponds to the initial phase difference of the two species of $\pi/2$. If the initial condition for the phase difference is not set to $\pi/2$, it will relax to $\pi/2$ with a relaxation time that is controlled by the wavevector of the solution, as shown in Ref.~\cite{pisegnaEmergentPolarOrder2024}.
Note that $\varphi_a(t, \bm x; \bm k)$ solves the full non-linear equations of motion and is only possible due to the rotational symmetry of these equations.
The solution is illustrated in \autoref{fig: solution 1}(a).
The plot on the left shows the fields in real space, in the direction parallel to the wave vector $\bm k$.
We may also visualize the configuration in $(\varphi_1, \varphi_2)$-space by following the line parametrized by $x$, shown on the right, where the minimum of the potential is indicated by dashed black lines.
This can be thought of as mapping the circle $x \in S^1$ into field space, with shorter wavelength solutions corresponding to the mapping winding around the origin more times.
From the discussion in the previous section, we may explicitly evaluate the entropy production for the solution given in Eq.~\eqref{solution 1}:
\begin{align}
     \dot \sigma = \frac{\alpha^2}{uD}  k^2 (|r| - k^2).
\end{align}

To establish the stability of this solution, we write the solution in a polar form and consider perturbations in the radial and tangential directions.
This is detailed in Appendix \ref{section: linear stability}.
After changing to a rotating frame of reference, we obtain a time- and space-independent system of equations, which can be used for a stability analysis. We find that to ensure stability the criterion $k^2 < \frac{1}{3} |r|$ needs to hold.
Therefore, for a sufficiently large system,
the longest wavelength solution is always stable.
The velocity of the wave is controlled by the non-reciprocal parameter, with $\alpha=0$ corresponding to continuous phase separation.
As $\alpha$ increases, the wave starts to move.
This traveling solution is an example of a non-equilibrium steady-state solution~\cite{pisegnaEmergentPolarOrder2024}.

Decreasing further below $r=0$ yields shorter wavelength solutions.
Minimizing the free energy density $\Eff = - \frac{3}{4} u A^2$ corresponds to maximizing the amplitude of the wave.
The global minimum is the state with the longest wavelength $k = 2 \pi / L$ where $L$ is the system size, while solutions with higher wave numbers are metastable states, at least for $\alpha = 0$.
Assuming $|r| \gg k^2$, the amplitude of the wave is $A \approx \sqrt{-r / u}$, corresponding to mapping the circle in field space around the bottom of the potential, as illustrated in \autoref{fig: solution 1}(a).
In the metastable states for which $k = 2 \pi n / L$ and $n > 1$ the circle is mapped $n$ times around the brim of the hat.
As $n$ increases, the amplitude of the wave decreases as the kinetic term of the free energy acts as a tension in the line.
This solution is closely related to the topological defects this model admits~\cite{ranaDefectSolutionsNonreciprocal2024}, as far away from the core of a defect, the system forms a traveling sine-wave.
The general analysis holds for small finite $\bar \varphi$, see \texttt{video\_2.mp4} in\cite{SupplementalMaterial} for further illustration.
When $|r| < k^2$, Eq.~\eqref{solution 1} no longer corresponds to a stationary solution of the equations of motion, neither stable, metastable, nor unstable, as the tension has become too high.

We may also write down an approximate solution for $\alpha = 0$ close to the minimum of the potential $\bar \varphi = \sqrt{-r / u}$,
\begin{subequations}\label{eq: solution 2}
\begin{align}
  \varphi_1(\bm x, t; \bm k) &= \delta \cos 2 \phi(\bm x; \bm k) - \bar \varphi, \\
  \varphi_2(\bm x, t; \bm k) &= 2 \sqrt{\bar\varphi\delta} \cos \phi(\bm x; \bm k), \\
  \phi = \bm k \cdot \bm x, \hspace{15mm}
  \varphi' &= \sqrt{|r| / u}, \hspace{15mm}
  \delta = \varphi' - \bar \varphi.
\end{align}
\end{subequations}
We show in Appendix \ref{section: approximate solution} that for $\delta \ll \varphi'$ and in the long wavelength limit, this is a good approximation of the stationary solution.
While the previous solution corresponds to mapping a line parallel to $\bm k$ around the brim of the hat, this solution can be visualized as mapping this same loop to a small sector of the brim.
This is illustrated in \autoref{fig: solution 1}(b), where the analytical solution is plotted in gray, together with the numerical solution and the deviation between them.
The relaxation of the system towards this solution is illustrated in \texttt{video\_3.mp4} \cite{SupplementalMaterial}.

Videos 2 and 3 are simulations of the NRCH model with an added noise term of the form $\sqrt{2 D} \, \bm \nabla \cdot \bm \xi_a(\bm x, t)$, where $\bm \xi_a$ represents Gaussian white noise of unit strength with zero mean. The noise-strength in the simulations is set to $D = 2 \times 10^{-4}$ in the appropriate non-dimensionalized form. The value is chosen to show the robustness of the observed behavior to weak noise, while still illustrating the deviation away from the analytical solution of \autoref{fig: solution 1}(b).
The solutions presented correspond to infinite systems or systems with periodic boundary conditions. We expect the bulk properties to hold for other boundary conditions for sufficiently large system sizes. The effect of confinement can be implemented by incorporating a lower bound on the values of the wavevector, as a first approximation.

Video 2 illustrates that a traveling wave state close to that of Eq.~\eqref{solution 1} exists even for $\bar \varphi \neq 0$.
The video also illustrates the possibility of topological defects in this system, which may occur due to the initial conditions or added noise. In one dimension, these appear as point defects, with traveling-wave domains between them.
When the system is initiated at random initial conditions and the coupling is relatively weak, topological defects appear and form an arrested structure that disrupts global order, while sufficiently strong non-reciprocal coupling suppresses topological defects and creates global (polar) order~\cite{ranaDefectSolutionsNonreciprocal2024}, which is robust with respect to noise, even in two dimensions~\cite{pisegnaEmergentPolarOrder2024}. While weak noise is not expected to fundamentally change the picture provided by the linear stability analysis as long as defects are suppressed, strong noise can lead to nontrivial qualitative changes in the behavior of the system. 
A further possible complication is the appearance of pseudo-resonances~\cite{trefethenHydrodynamicStabilityEigenvalues1993}.
Due to the existence of non-reciprocal interaction, the linear-stability matrix can be non-normal for certain parameter-values. This can have the effect that even for systems with eigenvalues where the real part is negative (such that all linear perturbations ultimately decay) some modes (in the pseudo-spectrum) feature transient growth, thereby significantly amplifying perturbations, albeit remaining finite in the long time limit~\cite{trefethenSpectraPseudospectraBehavior2005}.
We note that nonlinear non-reciprocal terms have been shown to lead to the emergence of effervescence, as well as self-generated noise due to spatiotemporal chaos~\cite{sahaEffervescenceBinaryMixture2025}.

\subsection{Relation to the Complex Ginzburg-Landau equation}

We would like to highlight that a non-conserved version of our $\mathrm{SO}(2)$ model, which can be written as
\begin{equation}
  \label{CGLE}
	\partial_t \varphi_a 
	= -
    \left[ 
        \pdv{\Ve_\mathrm{cs}}{\varphi_a} -\nabla^2 \varphi_a  + \alpha \epsilon_{ab}\varphi_b 
    \right],
\end{equation}
has similarities to the Complex Ginzburg-Landau equation (CGLE)~\cite{aransonWorldComplexGinzburgLandau2002}. One may consider CGLE as the non-reciprocal extension of model A~\cite{halperinCalculationDynamicCritical1972}, and the $\mathrm{SO}(2)$ NRCH, Eq.~\eqref{eom}, as the non-reciprocal extension of model B.
The CGLE may include higher-order non reciprocal terms, which have been explored recently for the NRCH~\cite{sahaEffervescenceBinaryMixture2025}.
In the case of the CGLE, these terms are necessary in order to obtain non-trivial behavior, as the linear term can be removed by a time-dependent field transformation, $\varphi_a \rightarrow [e^{- \alpha t \epsilon }]_{ab} \varphi_b$.
One might, naively, extend this argument to the $\mathrm{SO(2)}$ NRCH as well, with the transformation $\varphi_a \rightarrow [e^{\alpha \nabla^2 t \epsilon }]_{ab} \varphi_b$. Applying this transformation to the traveling-wave solution Eq.~\eqref{solution 1} gives a static-wave solution that solves the $\alpha = 0$ equation of motion. This would hint at a hidden $\Z_2$ symmetry, so that the full symmetry is $\mathrm{O}(2) \simeq \mathrm{SO}(2) \rtimes \Z_2.$
However, the $\varphi^2$ term is no longer invariant of the transformation when including the Laplacian operator in the exponential, so the transformation cannot be used to eliminate the linear non-reciprocal term.
Indeed, such a transformation would render the NRCH with only a linear non-reciprocal term more-or-less equivalent to model B. However, this model admits a broken-symmetry state in two dimensions~\cite{pisegnaEmergentPolarOrder2024}, which is not the case for model B. Furthermore, while Eq.~\eqref{CGLE} admits a Boltzmann distribution as a steady state, as one would expect since it can be mapped to an equilibrium theory, the same is \emph{not} the case for Eq.~\eqref{eom}.

A similar observation was made in Ref.~\cite{janssenRenormalizedFieldTheory1992}---the conserved version of the KPZ was at first believed to obey a modified version of the Galilean symmetry.
This symmetry is protects the KPZ non-linearity from renormalization, and is thus crucial for the existence of the KPZ universality class. However, more careful analysis shows that this is not true in the conserved case, and the non-linearity is indeed renormalized~\cite{janssenRenormalizedFieldTheory1992}, illustrating how careful analysis of symmetries are crucial in non-equilibrium models when considering conservation laws.

\section{Conclusion}

In this article, we have explored the phase behavior of the non-reciprocal Cahn-Hilliard model.
By choosing two reduced models---one with rotational invariance in field space and one without---we can fully characterize the linear behavior and see the effect of continuous symmetry on phase separation.
The main behaviors observed are homogeneous configurations, static phase separation---continuous or with a well-defined domain wall---and non-equilibrium steady states with non-zero currents.
Using this construction we have shown the effect of a continuous symmetry on phase separation, where there is no well-defined domain wall.
This effect remains in the case of traveling wave states where the phases chase each other due to non-reciprocal interactions.

In addition, we start to see a glimpse of the rich behavior close to the transition between static and traveling states, which linear analysis can not capture.
This complicated, chaotic seeming behavior has been shown in recent simulations of both particle systems~\cite{duanDynamicalPatternFormation2023} and the NRCH~\cite{sahaEffervescenceBinaryMixture2025}.
This is an area---the transition between regions dominated by the chasing behavior and the equilibrium behavior---which more sophisticated tools are needed to characterize.
We believe the results from such work will uncover interesting effects that further illuminate the complex behavior of active matter. The investigation will probe the frontier between two different but simple states, the static and traveling waves, and it is exactly near such boundaries that complexity tends to arise. We will relegate such studies to future work.

\begin{acknowledgements}
The Authors would like to thank Giulia Pisegna, Jacopo Romano, and Suropriya Saha for stimulating discussions. We acknowledge support from the Max Planck School Matter to Life and the MaxSynBio Consortium which are jointly funded by the Federal Ministry of Education and Research (BMBF) of Germany and the Max Planck Society.
\end{acknowledgements}

\appendix

\section{Linear stability of $\bar \varphi = 0$ solution}
\label{section: linear stability}

\begin{figure}[t]
  \centering
  \includegraphics[width=\columnwidth]{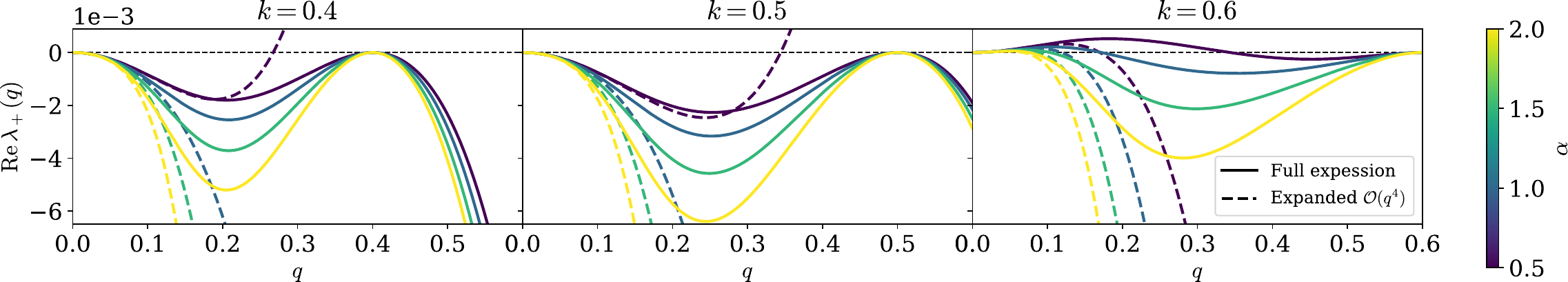}
  \caption{
    Eigenvalues for perturbation away from the traveling wave state, for different values of $k$, with $|r| = 1$ and $\bm k \cdot \bm q = k q$, for different values of $k = |\bm k|$ and $\alpha$.
    The rightmost panel corresponds to a system beyond the Eckhaus-instability $k = 1 / \sqrt{3}\approx 0.577$.
  }
  \label{fig: eigen}
\end{figure}

We now consider the stability of the traveling-wave solution of the $\mathrm{SO}(2)$ model.
We write it in polar form as $\varphi_a(\bm x, t) = A(\bm k) [e^{\epsilon \phi(\bm x, t; \bm k)}]_{ab} \hat \varphi_{b}$, were $\hat\varphi$ is a unit vector.
The solution in Eq.~\eqref{solution 1} corresponds to $\hat \varphi = (0, 1)^T$, while a rotation of $\hat\varphi$ corresponds to a phase shift.
The amplitude is given by $uA^2 = - r - k^2$, and $\phi = \bm k \cdot \bm x - \alpha k^2 t$.
The wave-vector $\bm k$ thus points in the direction of travel.
We then consider perturbations around this state, $A \rightarrow A + \rho$ and $\phi \rightarrow \phi + \theta$, to first order in the perturbations, 
We can find the stability of the solution by considering 
\begin{align}
    \varphi_a(\bm x, t) & = 
    [A(\bm k) + \rho(\bm x, t)]
    [e^{\epsilon[\phi(\bm x, t; \bm k) + \theta(\bm x, t)] }]_{ab} \hat \varphi_a,\\
    & = 
    [A(\bm k) + \rho(\bm x, t)]
    [\cos\{\phi(\bm x, t; \bm k) + \theta(\bm x, t)\} \delta_{ab} + \sin\{\phi(\bm x, t; \bm k) + \theta(\bm x, t)\}\epsilon_{ab}] \hat \varphi_a.
\end{align}
We insert this into the equation of motion Eq.~\eqref{eom}, which gives
\begin{align}
    &
	[ \partial_t \rho + \epsilon(-\alpha k^2 \rho_0 - \alpha k^2 \rho + \rho_0 \partial_t \theta) ]
    e^{\epsilon(\phi + \theta)} \hat \varphi
	= 
	\big\{
		[(r + u\rho_0^2)  + \alpha \epsilon]  
		[- \rho_0 k^2 + (a_2 + \epsilon b_2) \rho + (c_2 + \epsilon d_2)\theta],
		\\ &  \hspace{18mm}
		+ 
		2 u \rho_0
		[
		\rho_0 (\nabla^2 - k^2) \rho
		+
		\epsilon (2 \rho_0 \bm k \cdot \bm \nabla) \rho
		]
		- \rho_0 k^4
		+
		(a_4 + \epsilon b_4) \rho + (c_4 + \epsilon d_4)\theta
	\big\}
    e^{\epsilon(\phi + \theta)} \hat \varphi ,
\end{align}
to leading order, where the derivative operators 
\begin{align}
	a_2 & = \nabla^2 - k^2,
    & a_4 & = -[(\nabla^2 - k^2)^2 - 4 (\bm k \cdot \bm \nabla)^2], \\
    b_2 & = 2 \bm k \cdot \bm  \nabla ,
    & b_4 & = -[4(\nabla^2 - k^2) \bm k \cdot \bm \nabla ], \\
	c_2 & = - 2 \rho_0 \bm k \cdot \bm \nabla,
    &c_4 & = \rho_0[4(\nabla^2 - k^2) \bm k \cdot \bm \nabla ], \\
	d_2 & = \rho_0 \nabla^2,
	& d_4 & = - \rho_0\left[(\nabla^2 - 2k^2)\nabla^2 - 4 (\bm k \cdot \bm \nabla)^2\right],
\end{align}
only affect $\rho$ and $\theta$ within the parenthesis, not the exponential functions.
With this form, we can change to a rotating frame of reference where we write a set of two linear equations for the perturbations as follows
\begin{align}
	\partial_t \rho &=
	[ -k^2 a_2 - \alpha b_2 - 2(r + k^2)  (\nabla^2 - k^2) + a_4 ] \rho 
	+ [-k^2c_2 - \alpha d_2 + c_4  ] \theta,\\
	\rho_0 \partial_t \theta
	& =  
	[ \alpha a_2 - k^2 b_2 + \alpha k^2 - 4u(r + k^2)\bm k \cdot \bm \nabla + b_4 ]\rho
	+
	[\alpha c_2 -k^2 d_2 + d_4]\theta.
\end{align}
Note that all coefficients here are independent of space and time.
The stability of the solution is then given by the eigenvalues of the corresponding matrix in Fourier space
\begin{align}
	& \lambda_{\pm} (\bm q; \bm k)
    = 
    \\\nonumber & \quad
	k^4-|r|  \left(k^2+q^2\right) -4 ({\bm k \cdot \bm q})^2+2 i \alpha  ({\bm k \cdot \bm q})-q^4
	\pm \big\{k^8+k^4 \left(-4 ({\bm k \cdot \bm q})^2+q^4-4 q^2 |r| +|r| ^2\right) 
    \\\nonumber & \quad
    +2 k^6 \left(q^2-|r| \right) 
    +2 k^2 |r|  \left(4 ({\bm k \cdot \bm q})^2-q^4+q^2 |r| \right)+q^4 (4 ({\bm k \cdot \bm q})-i \alpha )^2+4 ({\bm k \cdot \bm q}) q^2 |r|  (4 ({\bm k \cdot \bm q})-i \alpha )+q^4 |r| ^2\big\}^{\frac{1}{2}}.
\end{align}
Here, the condition $\mathrm{Re} (\lambda) > 0$ implies that the corresponding eigen-mode grows exponentially.
The most unstable eigenvalue is $\lambda_+$, which corresponds to perturbations parallel to the traveling wave ($\bm k \cdot \bm q =  k q$). This mode is illustrated with solid lines in Fig.~\ref{fig: eigen}, for different values of $k$ and $\alpha$.
For plots of orthogonal perturbations, $\bm k \cdot \bm q = 0$, and the imaginary part, see the Supplemental Material~\cite{SupplementalMaterial}.
We note that the real part has a zero when the wave-vector of the perturbation matches the traveling wave, i.e. $\lambda(\bm q = \bm k; \bm k)  = i \alpha k^2$, which corresponds to a phase-shift.
The $-\nabla^4$ ensures the stability against high wave-number perturbations.
We set $\bm q \cdot \bm k = q k$, and expand to obtain
\begin{align}
    \lambda_{+,\parallel}(\bm q; \bm k)
    & = 2i \alpha k q - D_{2, \parallel}(k) q^2 + i \Omega_{\parallel}(k) q^3 - D_{4, \parallel}(k) q^4 + \Oh(q^5).
\end{align}
This approximate form is plotted as dashed lines in Fig.~\ref{fig: eigen}, and compared with solid lines that correspond to the full expression. The important parameter here is
\smash{$
    D_{2, \parallel} = \frac{|r| - 3k^2}{|r| - k^2} k^2,
$}
which changes sign at the onset of the Eckhaus instability, thus requiring $k^2 < \frac{1}{3}|r|$ as the condition for stability~\cite{aransonWorldComplexGinzburgLandau2002,pisegnaEmergentPolarOrder2024}.
This shows that the long wavelength solution is always stable.
For a finite system of linear dimension $L$, there may exist solutions where $k = 2 \pi n / L$ for $n>1$, corresponding to wrapping a line in the domain off $\varphi_a(\bm x)$ several times around the center of the potential, but in the $\alpha = 0$ case, at least, this is a metastable state.

\section{Approximate solution}
\label{section: approximate solution}

We now insert the configurations Eq. \eqref{eq: solution 2} into the equations of motion of the field.
We have
\begin{align}
  \varphi^2 & = \delta^2 \cos^22\phi 
  -2 \bar \varphi \delta {\cos2\phi} + \bar\varphi^2 + 4 \bar\varphi\delta\cos^2\phi,\\ 
  & = \bar\varphi^2 - 2 \bar\varphi \delta + \Oh(\delta^2).
\end{align}
The left-hand sides of the equation of motion vanish by assumption, as the configurations are time-independent, while the right-hand sides are
\begin{align}
	\Gamma \nabla^2
	\left\{
		\left[r - \nabla^2 + u \left(\bar \varphi^2 + 2 \bar \varphi \delta\right)\right]
	\right\} (\delta \cos 2\phi - \bar \varphi) + \Oh(\delta^3)
	= - 4 \delta \Gamma k^2 \left[
		r + 4 k^2 + u\left(\bar \varphi^2 + 2 \bar \varphi \delta\right)
	\right] \cos 2\phi + \Oh(\delta^3),\\
	\Gamma \nabla^2
	\left\{
		\left[r - \nabla^2 + u \left(\bar \varphi^2 + 2 \bar \varphi \delta\right)\right]
	\right\} \sqrt{\bar\varphi\delta}\cos\phi + \Oh(\delta^{5/2})
	= - 2 \sqrt{\bar\varphi\delta} \Gamma k^2 
	\left[
		r + 4 k^2 + u\left(\bar \varphi^2 + 2 \bar \varphi \delta\right)
	\right] \cos \phi + \Oh(\delta^{5/2}).
\end{align}
We now use $\bar\varphi^2 = {\varphi^*}^2 + \Oh(\delta) = - r/u + \Oh(\delta) $ to rewrite
$r + u(\bar\varphi^2 + 2\bar\varphi\delta) = - 2 \bar \varphi \delta.$
Thus, when we let $\delta$, $k^2 \rightarrow 0$, the right-hand side of the equations of motion vanish as $\Oh(\delta^2 k^2, \, \delta k^4)$ and $\Oh(\delta^{3/2} k^2, \, \delta^{1/2} k^4)$, which are faster than the amplitudes of the solutions, ensuring convergence.

\section{Numerical methods}
\label{section: numerical methods}

The simulations are performed using finite difference methods in one dimension, giving a discretized set of coupled stochastic ODEs,
\begin{align}
	\partial_t \varphi_{a,n} &= \nabla^2_{nm} \Gamma \mu_{a,m} + \sqrt{\frac{2 D}{\Delta x}} \nabla_{nm} \xi_{a,m},
  \end{align}
on which apply the Euler-Maruyama method for time-stepping.
To guarantee detailed balance in the equilibrium ($\alpha = 0$) case, also in the space-discretized equations, it is important that the discretized Laplacian is exactly the square of the gradient acting on the noise, i.e., $\nabla^{2}_{nm} = \nabla_{nk} \nabla_{km}$~\cite{nardiniEntropyProductionField2017}---this is the Einstein relation.
Here $\xi$ is unit white noise, 
and we use the second-order stencil for $\nabla_{nm}$, 
\begin{align}
	\left( \frac{1}{12} , - \frac{2}{3}, 0, \frac{2}{3}, - \frac{1}{12}\right)
\end{align}
This has the disadvantage that high-frequency oscillations are not detected by $\nabla^2$ as it only measures sites at an even number of lattice sites away from the diagonal.
We therefore apply a different operator, $\tilde \nabla^2$ with the stencil 
\begin{align}
	\left( - \frac{1}{12} , \frac{4}{3}, - \frac{5}{2}, \frac{4}{3}, - \frac{1}{12}\right)
\end{align}
to the surface tension term, which becomes $\nabla^2 \tilde \nabla^2 \varphi$.
This ensures that high-frequency oscillations have an energetic cost and avoid a decoupling of the even and odd lattice sites.

\bibliography{ref,ref_manual}

\end{document}


\title{State Diagram of the Non-Reciprocal Cahn-Hilliard Model and the Effects of Symmetry \\ {\it Supplementary Material}}
\author{Martin Kjøllesdal Johnsrud\,\orcidlink{0000-0001-8460-7149}}
\affiliation{Max Planck Institute for Dynamics and Self-Organization (MPI-DS), D-37077 Göttingen, Germany}
\author{Ramin Golestanian\,\orcidlink{0000-0002-3149-4002}}
\email{ramin.golestanian@ds.mpg.de}
\affiliation{Max Planck Institute for Dynamics and Self-Organization (MPI-DS), D-37077 Göttingen, Germany}
\affiliation{Rudolf Peierls Centre for Theoretical Physics, University of Oxford, Oxford OX1 3PU, United Kingdom}
\date{\today}

\maketitle

\setcounter{equation}{0}
\setcounter{figure}{0}
\renewcommand{\theequation}{S\arabic{equation}}
\renewcommand{\thefigure}{S\arabic{figure}}

\section{Additional figures}

\begin{figure}[h]
  \centering
  \includegraphics[width=.375\columnwidth]{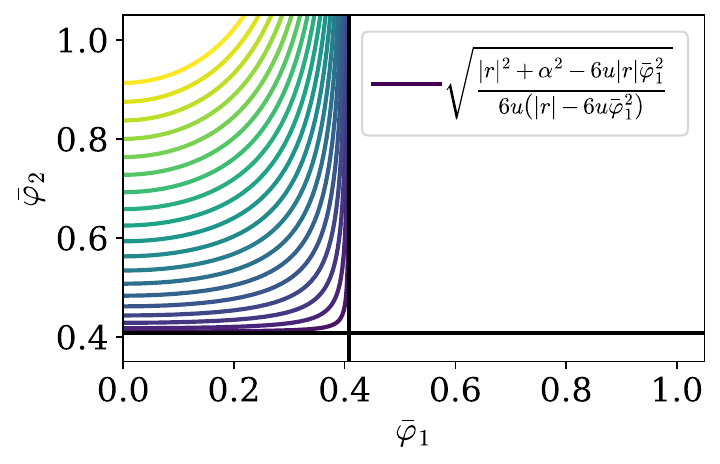}
  \caption{The critical line for the discrete model for $\alpha/|r| \in[0, 2]$. Brighter colors in the plot indicate higher values of $\alpha$.
  }
  \label{fig: critical assym}
\end{figure}

\begin{figure}[!htb]
	\centering
	\includegraphics[width=.88\textwidth]{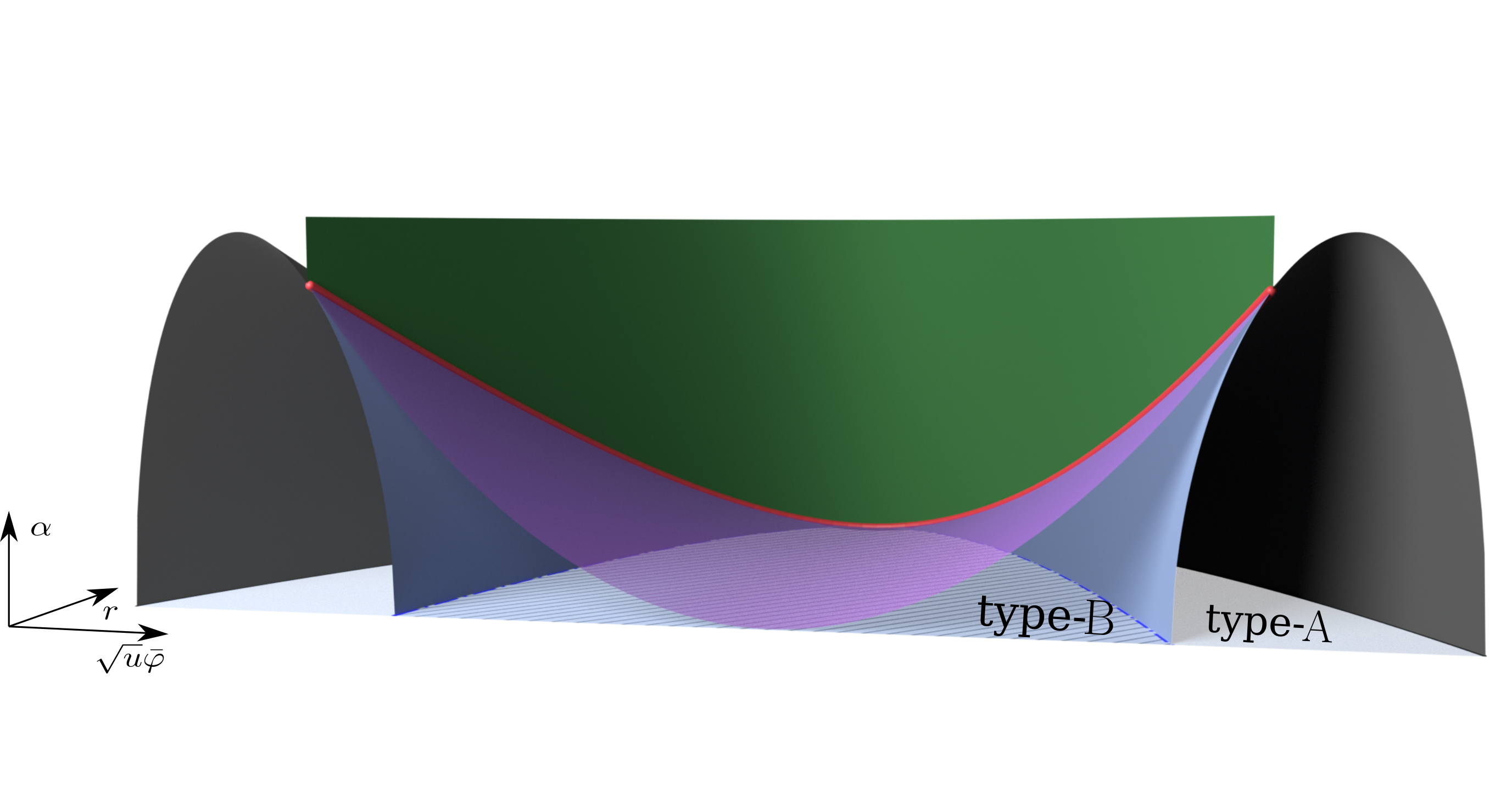}
	\includegraphics[width=.82\textwidth]{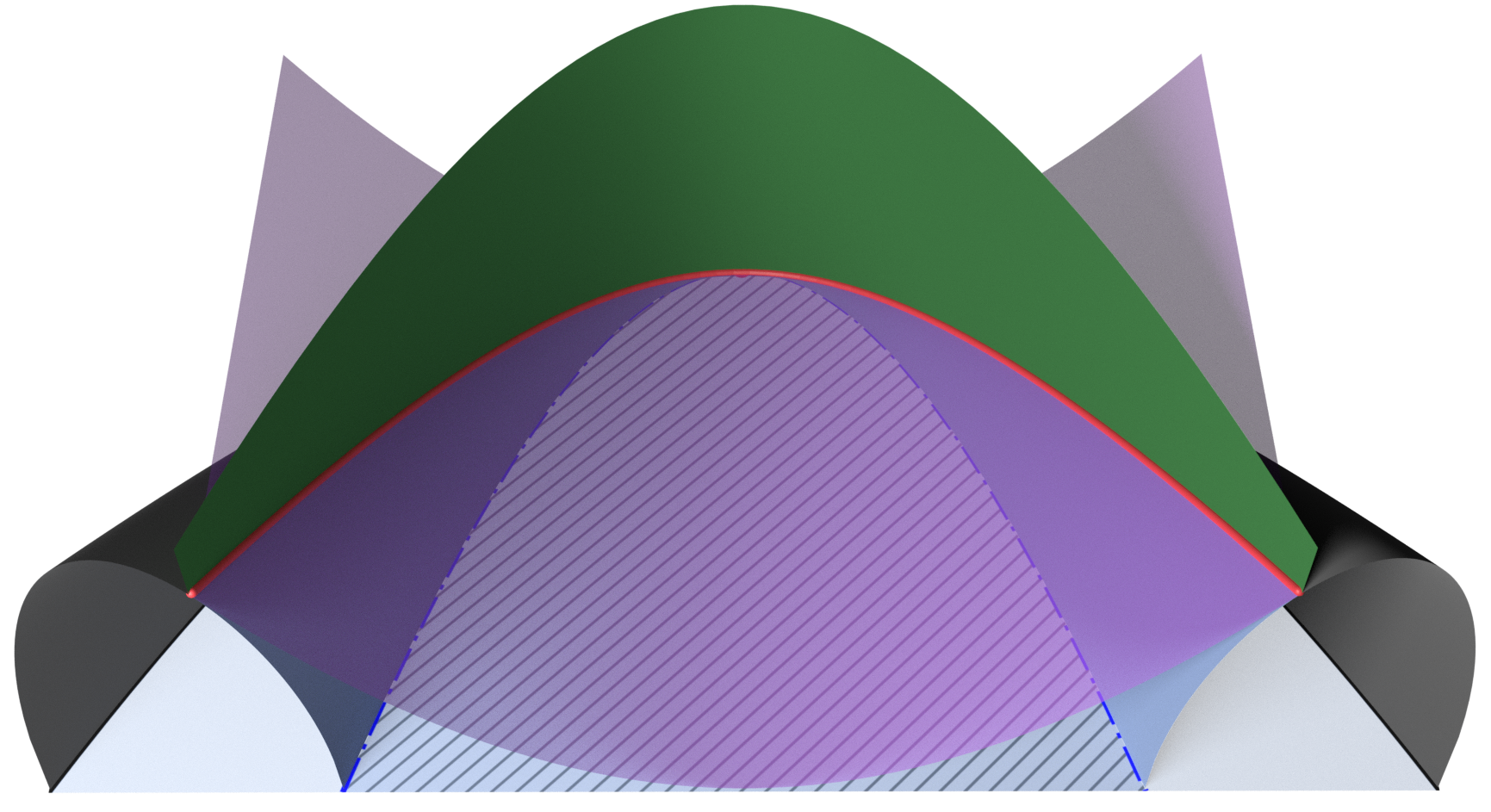}
	\includegraphics[width=.82\textwidth]{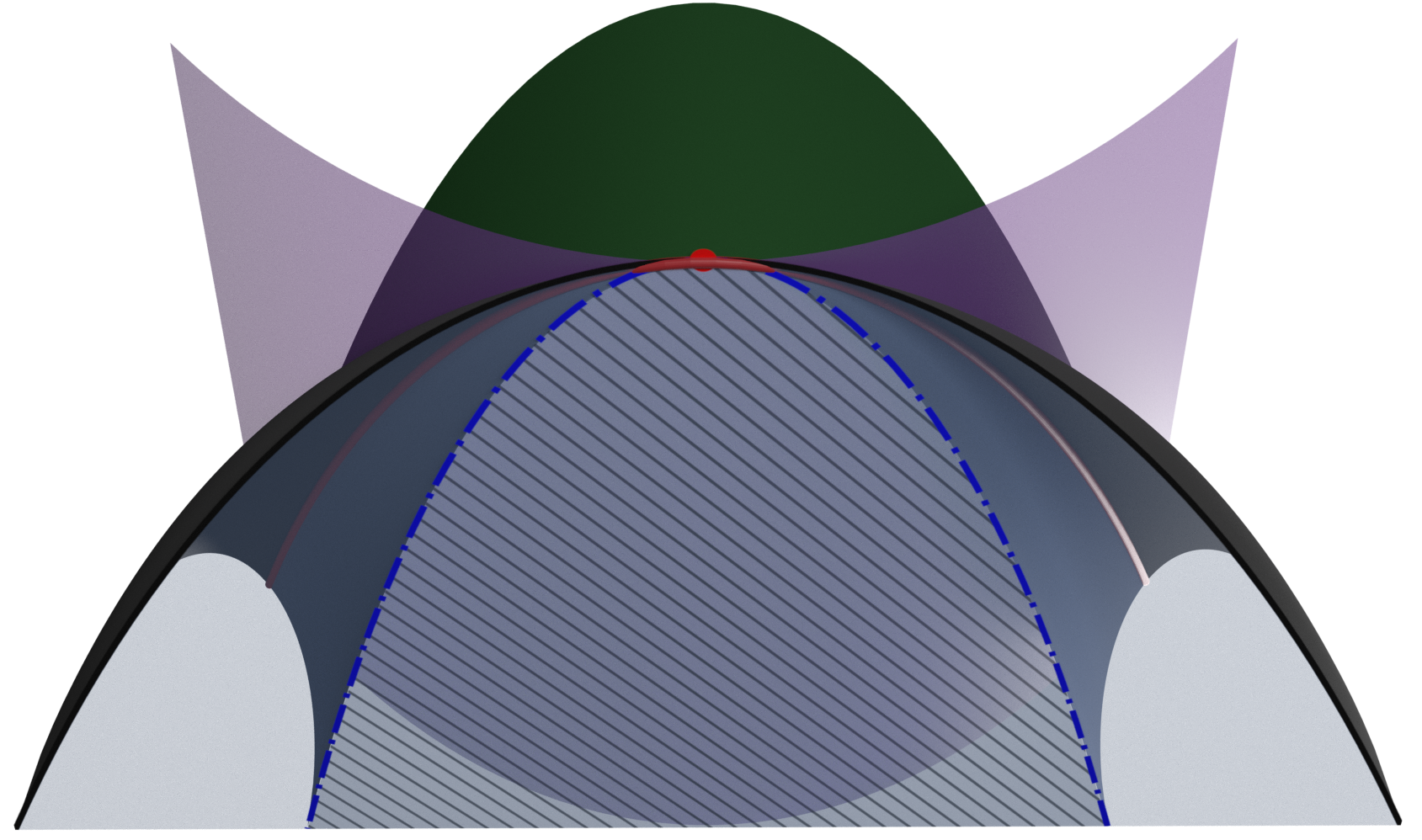}
	\caption{
	  The linear stability state diagram of the $\text{SO}(2)$ model.
	  }
	\label{fig: phase surface 1}
\end{figure}

\begin{figure}[!htb]
	\centering
	\includegraphics[width=.65\textwidth]{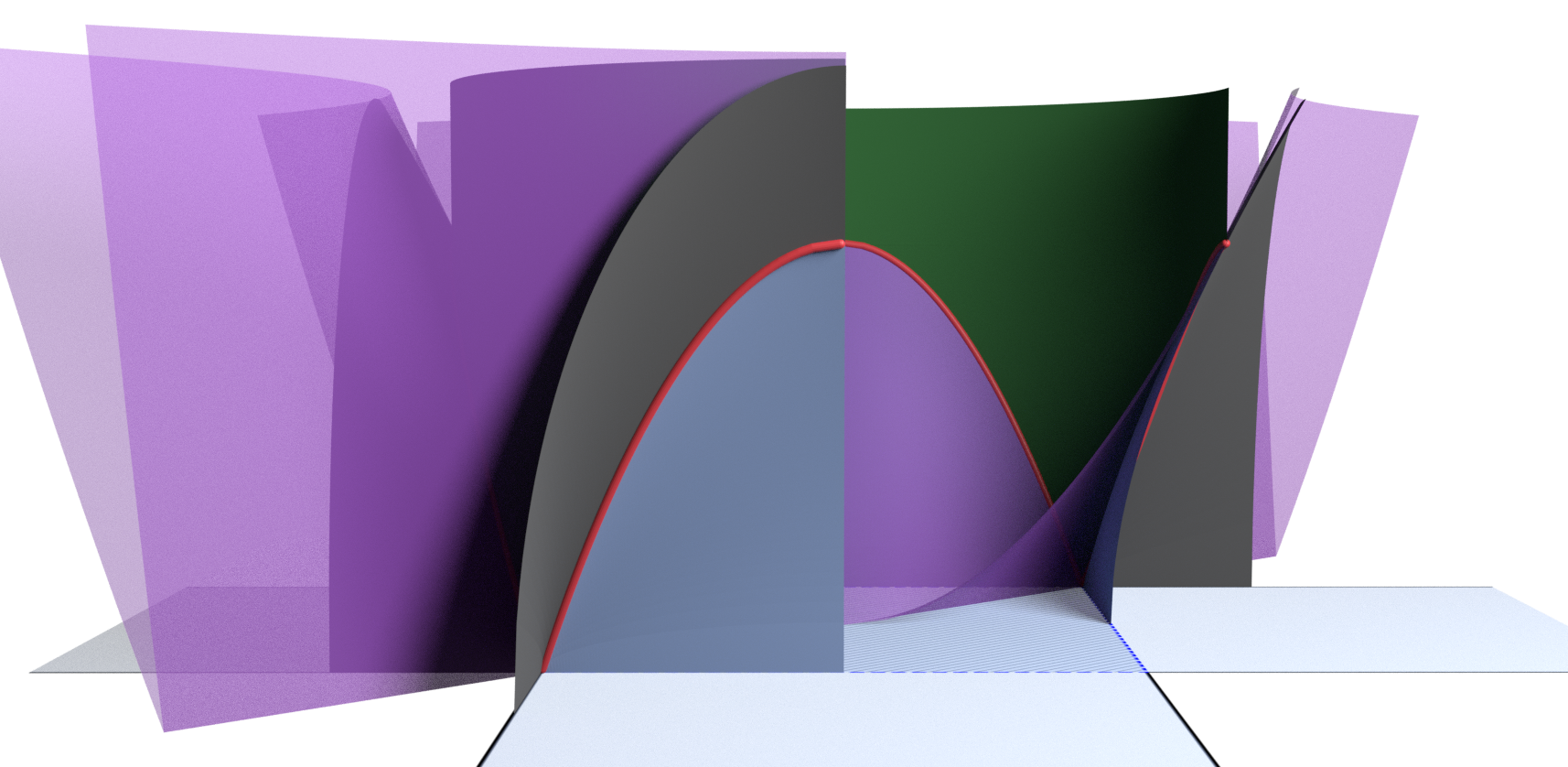}
	\includegraphics[width=.65\textwidth]{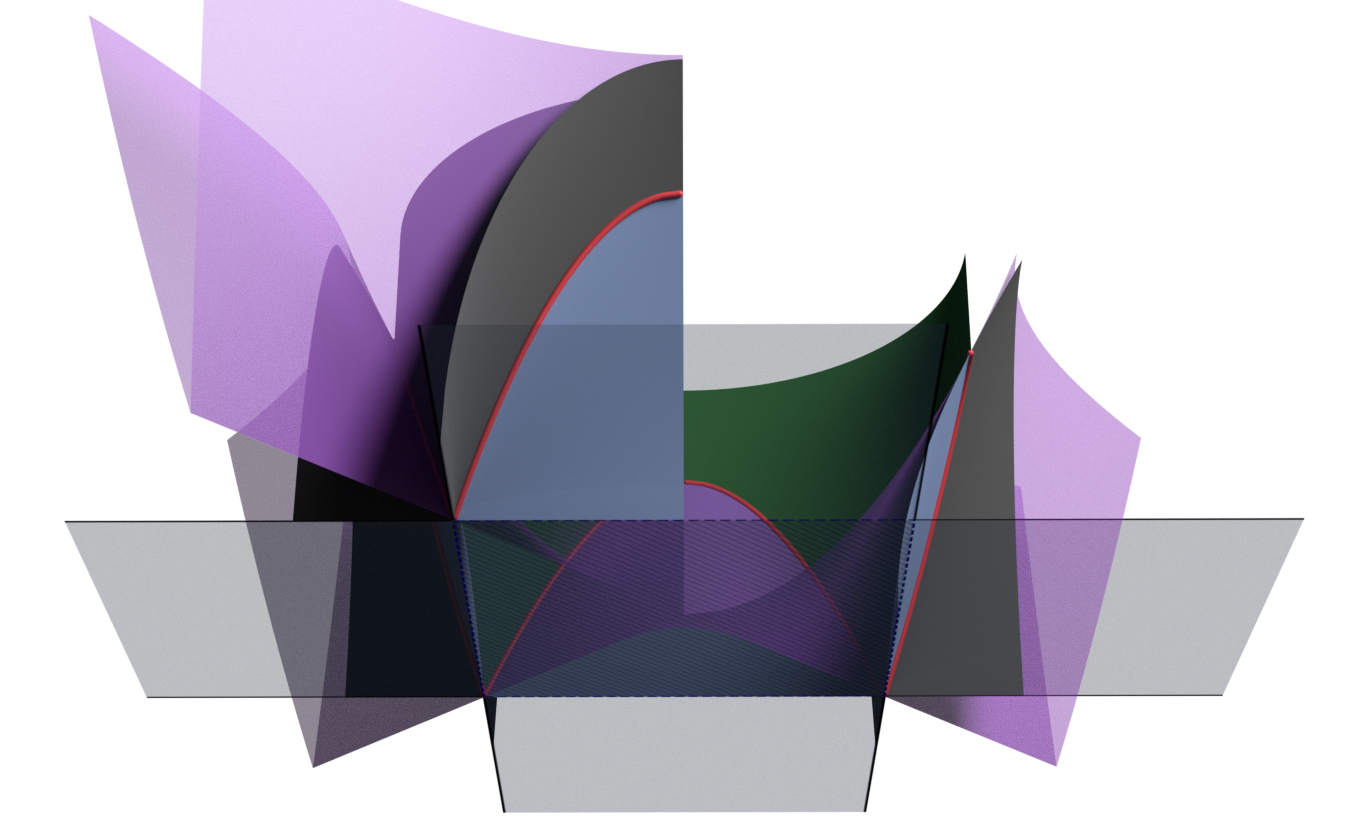}
	\includegraphics[width=.55\textwidth]{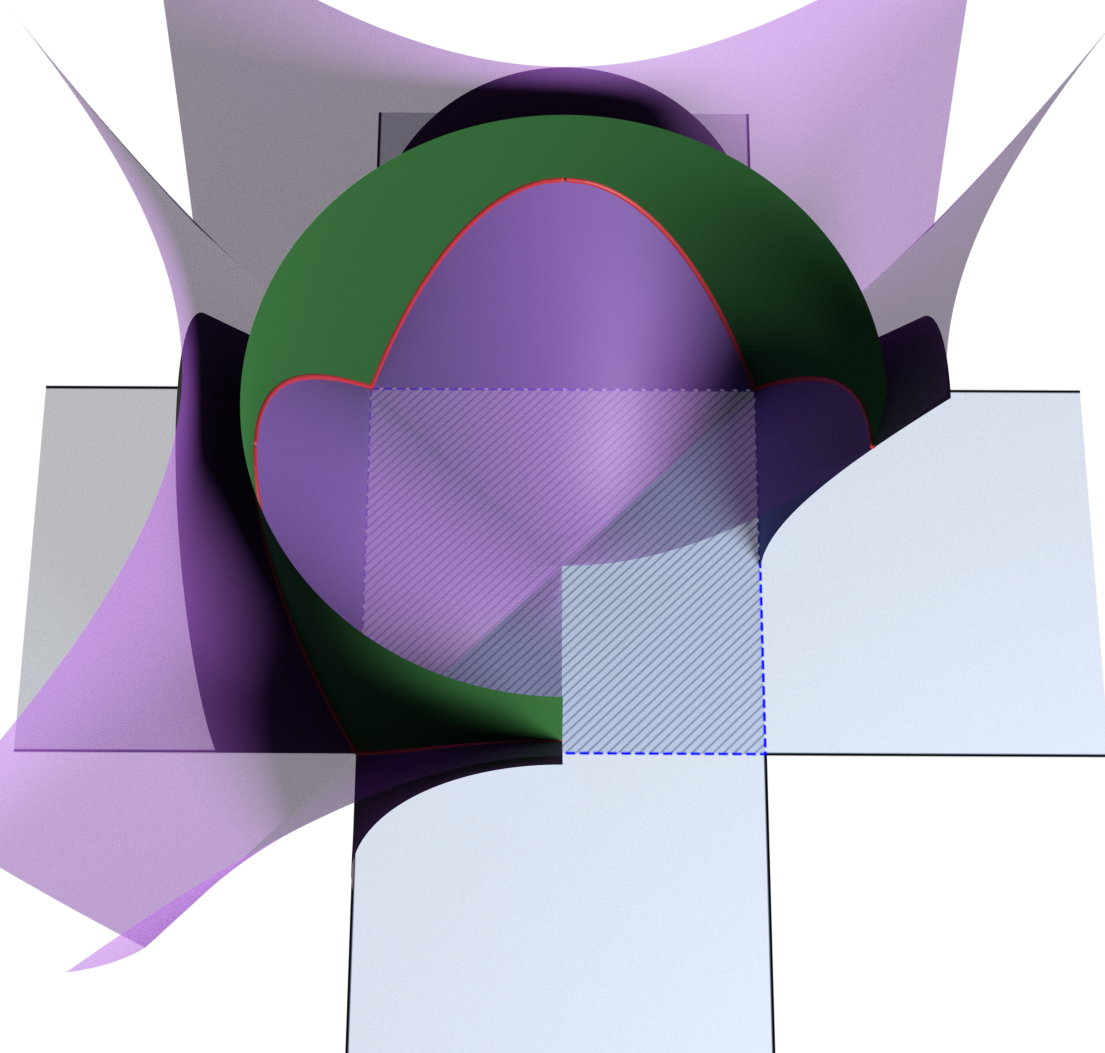}
	\caption{
	  The linear stability state diagram of the $\text{C}_4$ model.
	  }
	\label{fig: phase surface 2}
\end{figure}

\begin{figure}[!htb]
    \centering
    \includegraphics[width=.49\textwidth]{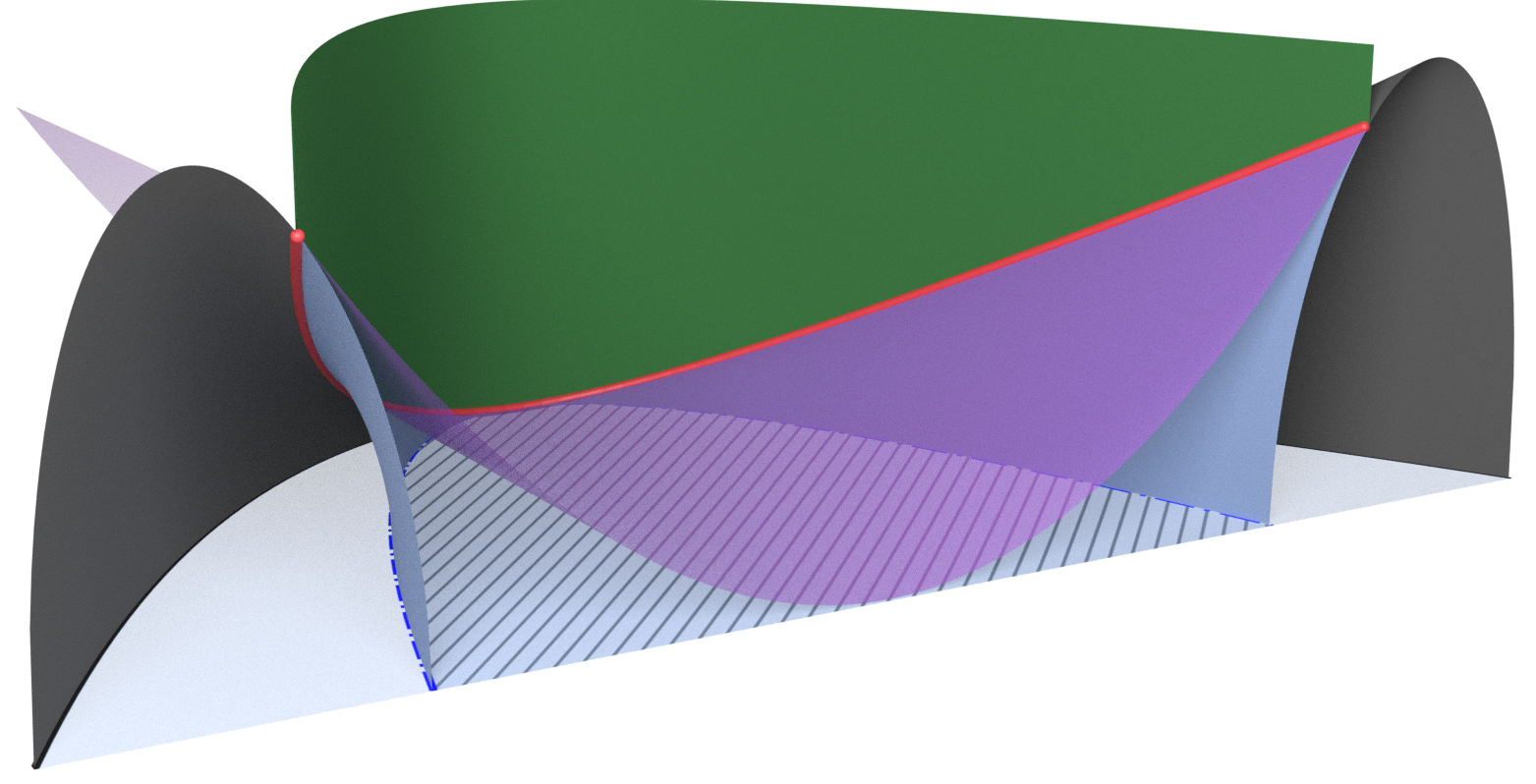}
    \includegraphics[width=.49\textwidth]{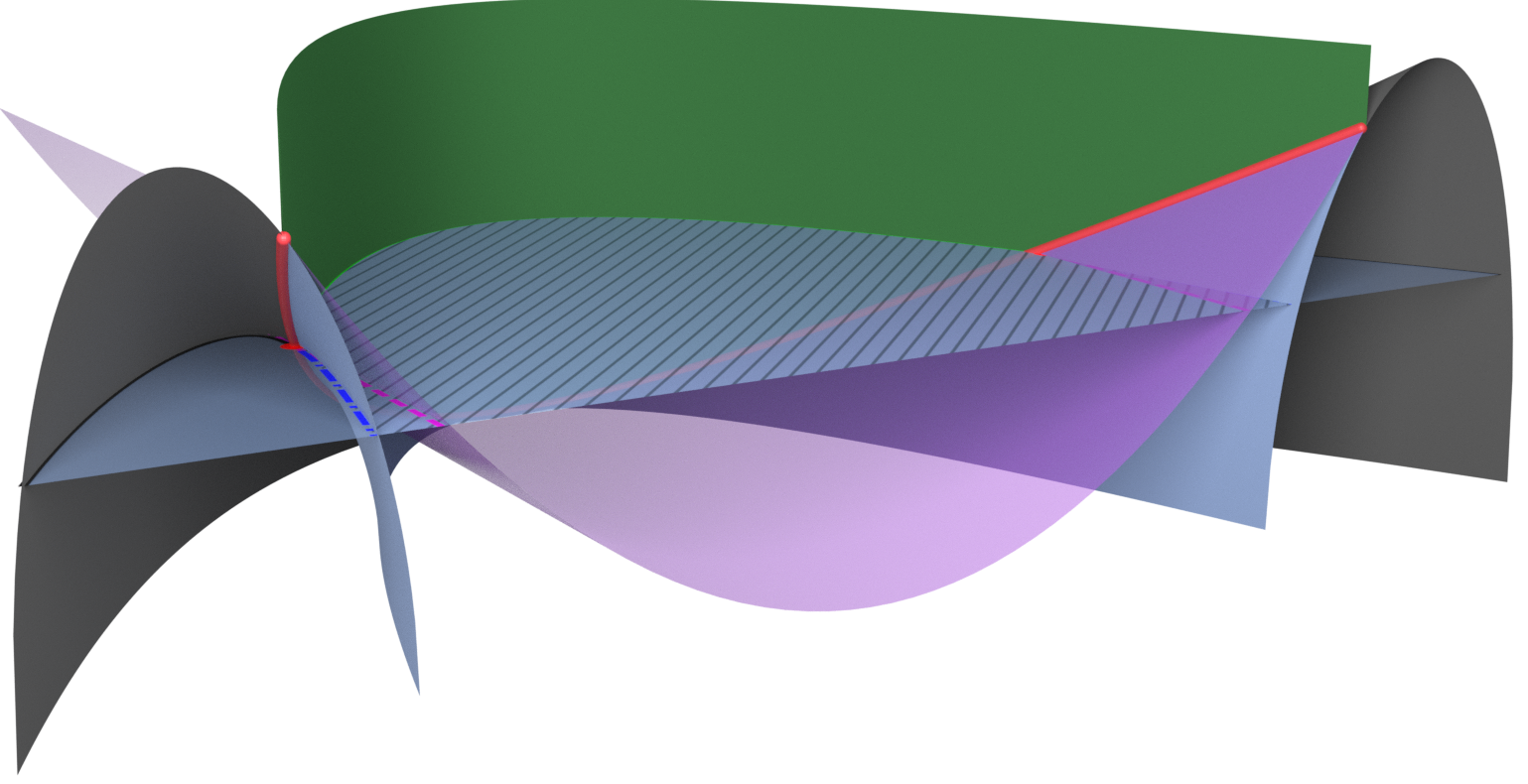}
    \includegraphics[width=.49\textwidth]{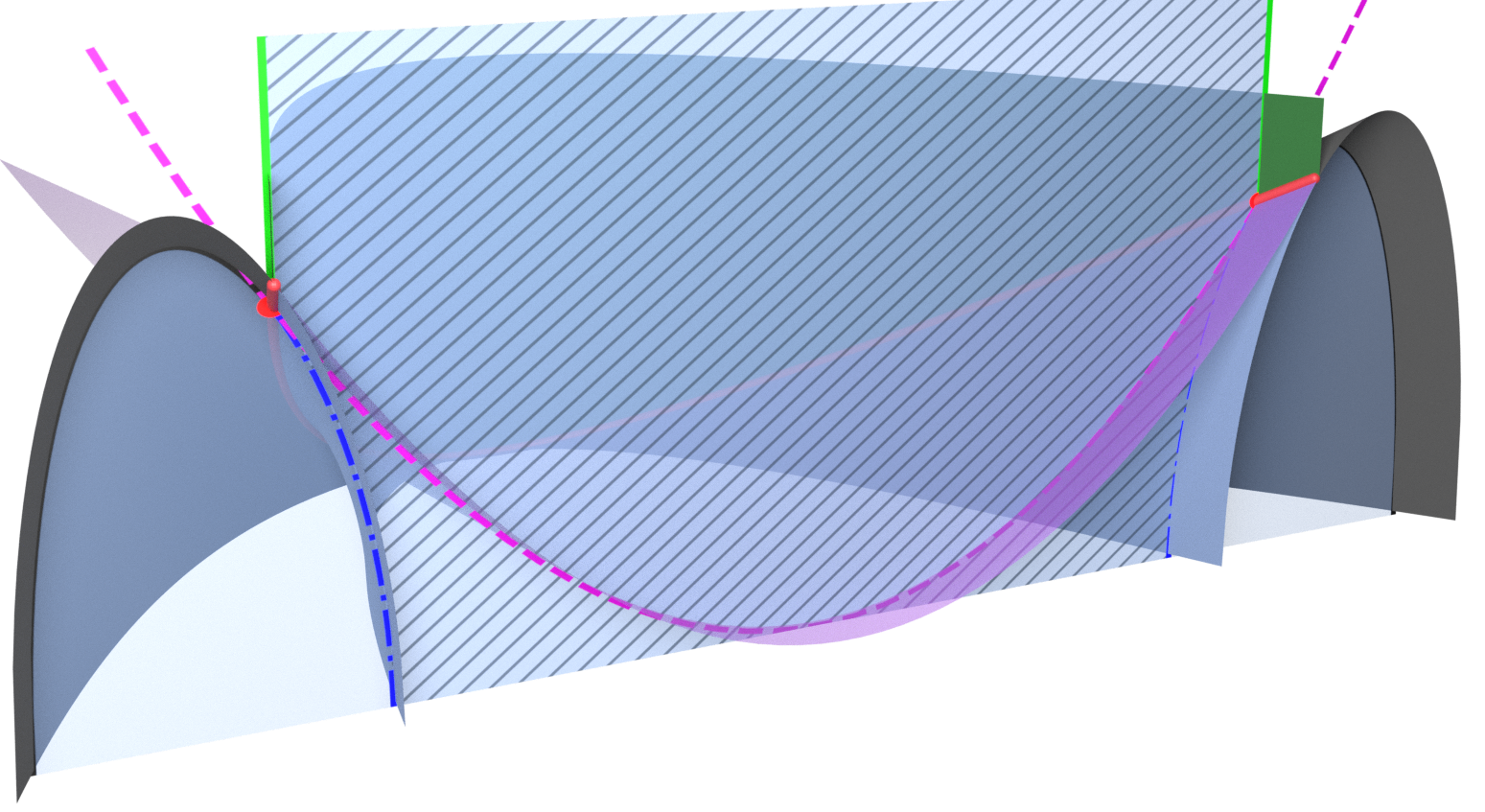}
    \includegraphics[width=.49\textwidth]{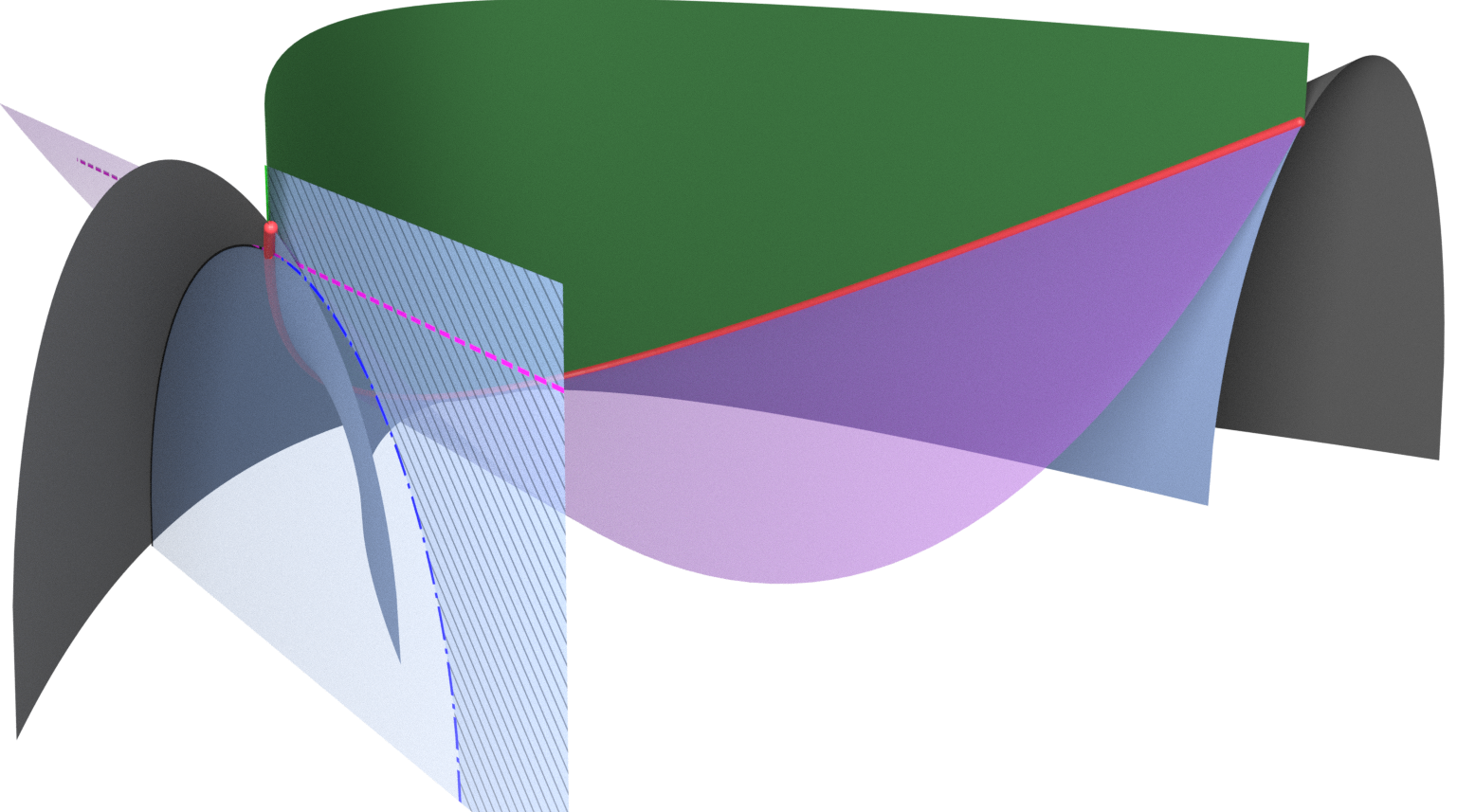}
	\caption{The cuts into the linear stability state diagram are placed in the full 3-dimensional diagram of the $\text{SO}(3)$ model.}
	\label{fig: inserts}
\end{figure}

\begin{figure}[!htb]
\centering
\includegraphics[width=.246\textwidth]{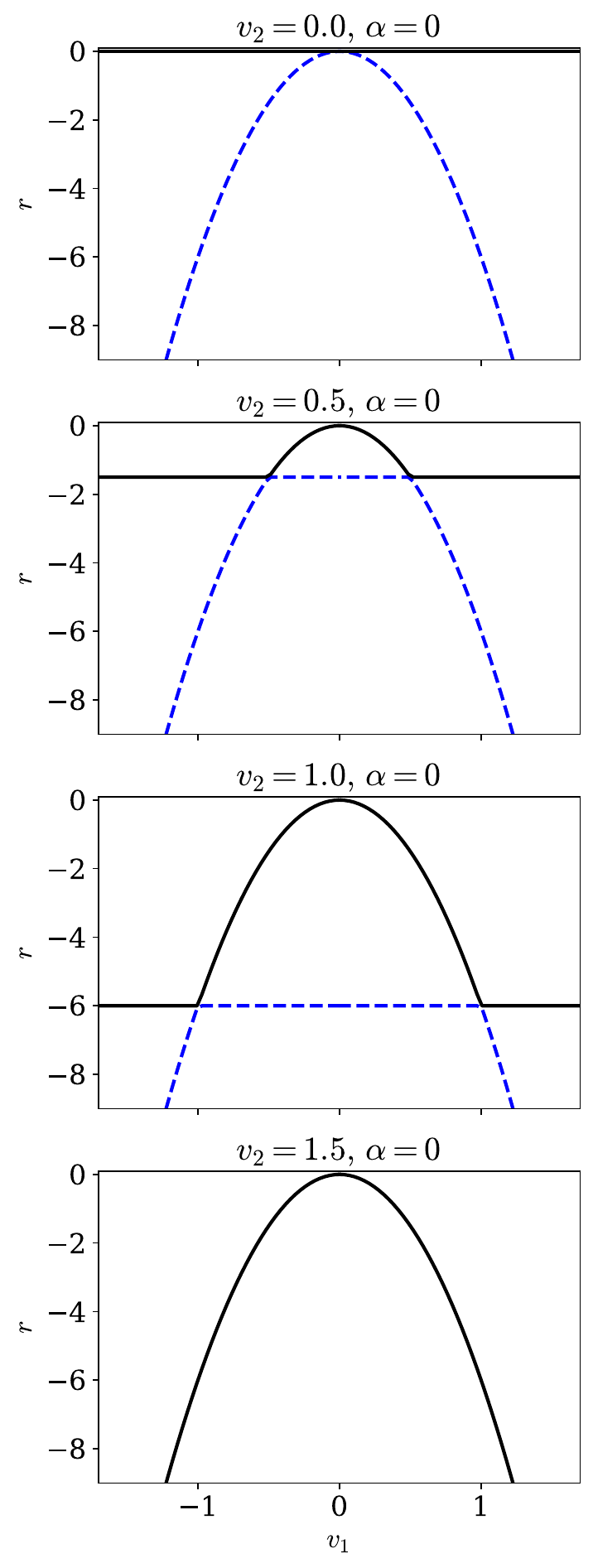}
\includegraphics[width=.25\textwidth]{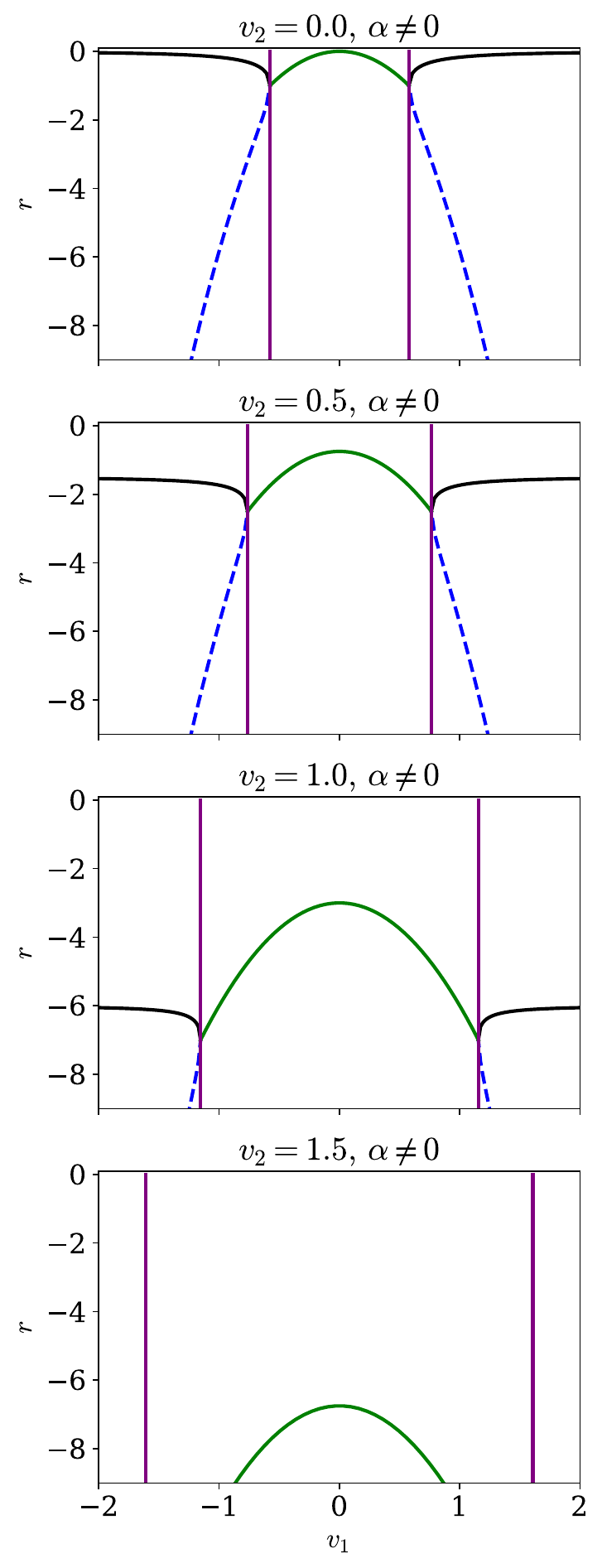}
\includegraphics[width=.24\textwidth]{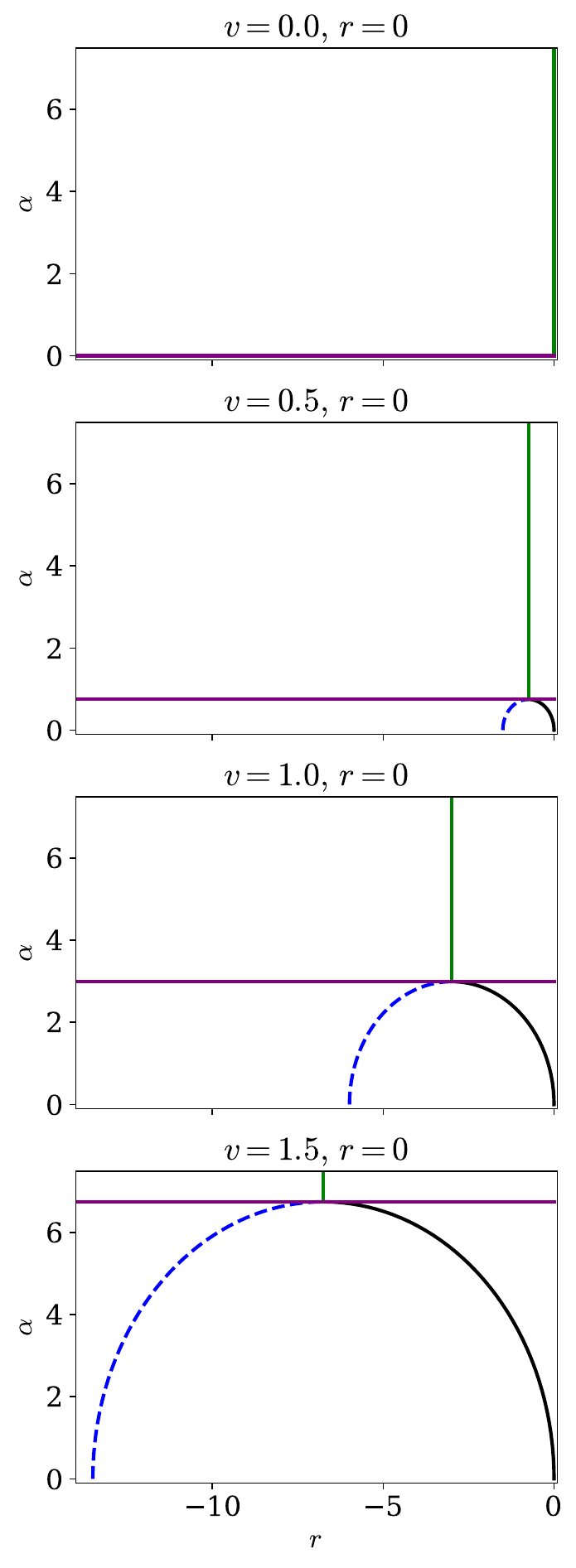}
\includegraphics[width=.241\textwidth]{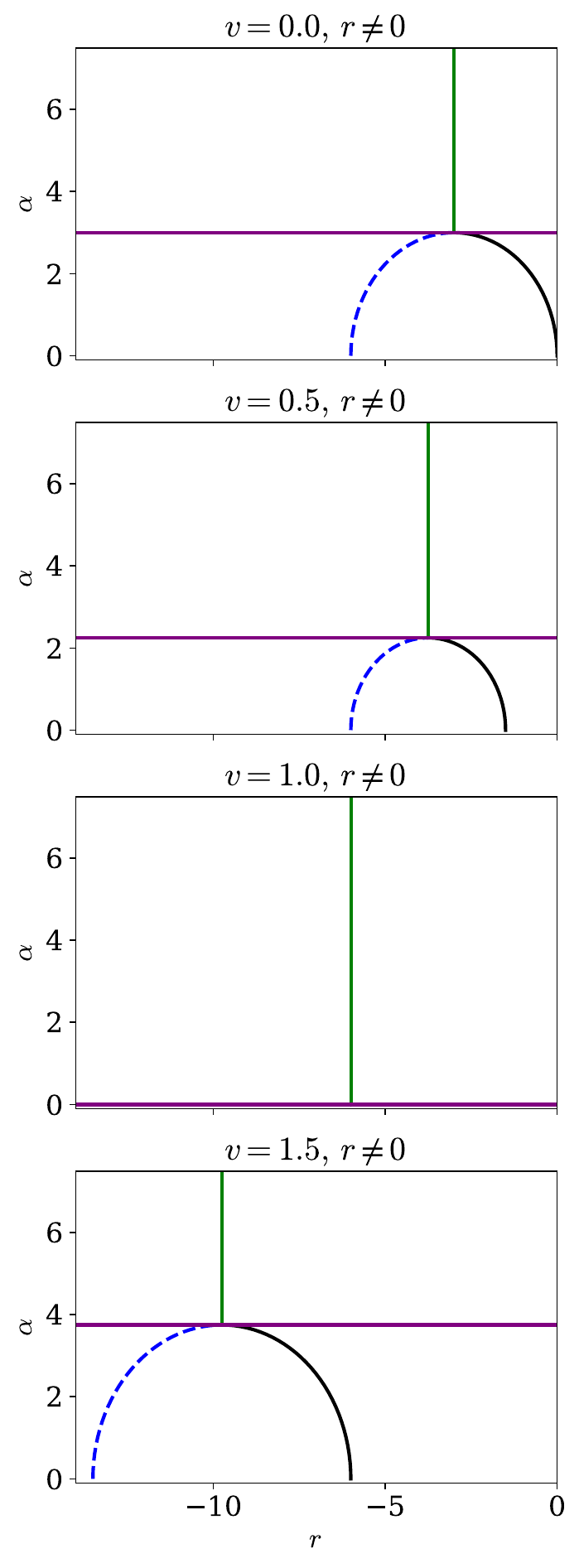}
\caption{
	The linear stability state diagrams, scaled by $\alpha$ (left) and $v_2$ (right).
	}
\label{fig: more C4 diagrams}
\end{figure}

\begin{figure}
    \centering
    \includegraphics[width=\textwidth]{fig/eigen.pdf}
    \includegraphics[width=\textwidth]{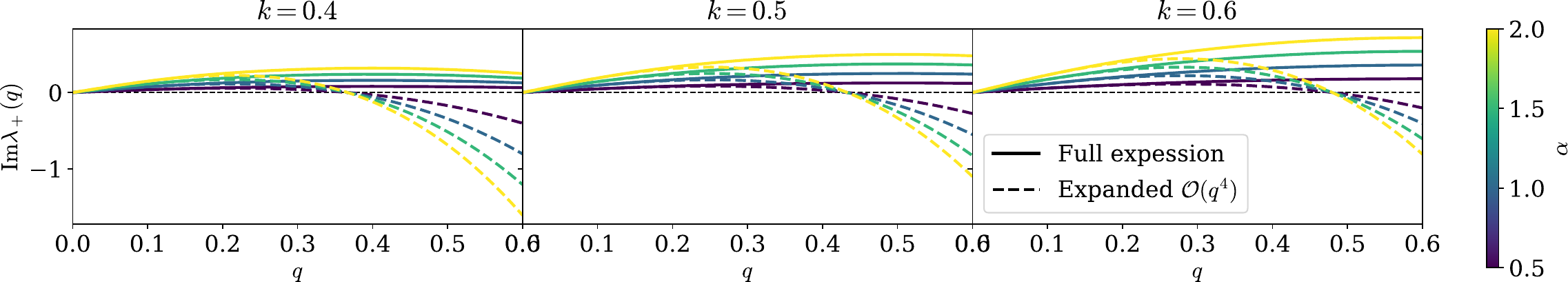}
    \includegraphics[width=\textwidth]{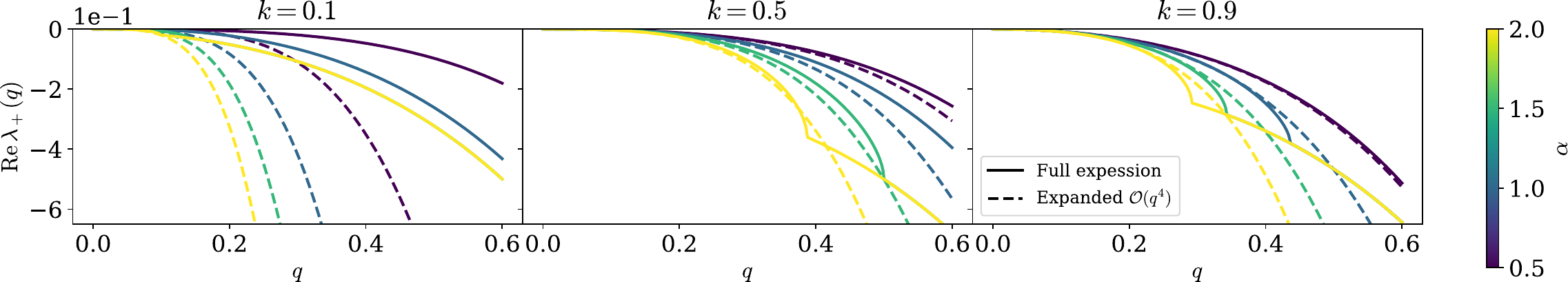}
    \includegraphics[width=\textwidth]{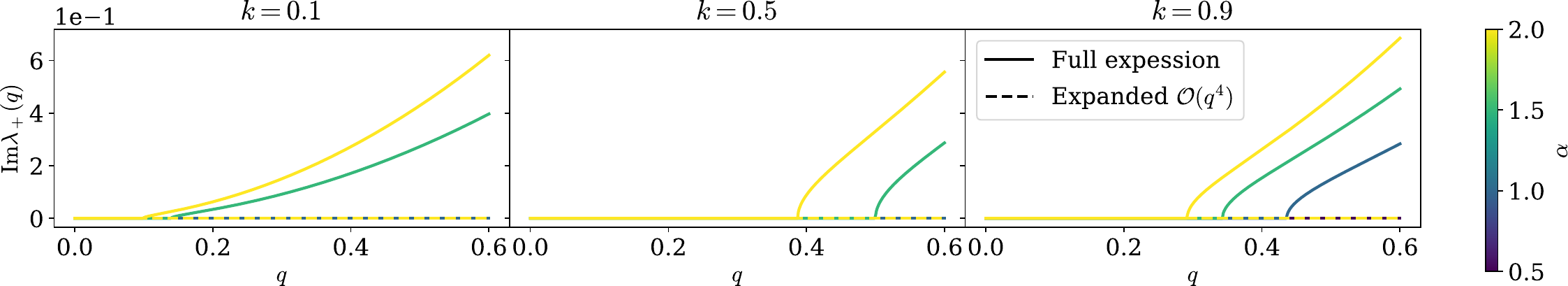}
    \caption{
    The real and imaginary part of the eigenvalue $\lambda_+(\bm q; \bm k)$ of perturbations of the traveling wave solution of the $\mathrm{SO}(2)$ system.  
    Full lines are the full expression, while dashed lines are expanded to fourth order in $q$.
    From top to bottom: $\mathrm{Re}\lambda_+(\bm k\cdot \bm q = qk)$, $\mathrm{Im}\lambda_+(\bm k\cdot \bm q = qk)$, $\mathrm{Re}\lambda_+(\bm k\cdot \bm q = 0)$,  and $\mathrm{Im}\lambda_+(\bm k\cdot \bm q = 0)$. In all cases, $r = -1$.
    We see that only the top-right case has positive real values, yielding an instability.
    }
    \label{fig:eigen}
\end{figure}

\autoref{fig: critical assym} shows the critical line for the discrete model.
We have included pictures of the full 3D state diagrams from additional angles.
\autoref{fig: phase surface 1} shows the $\text{SO}(2)$ model, and \autoref{fig: phase surface 2} the $\text{C}_4$ model.
To illustrate where the cuts in Figure 2 in the main text are done, we include inserting them into the 3D.
This is shown in \autoref{fig: inserts}.
The subplots are arranged as in the main text.
In the case of the $\mathrm{C_4}$ model, there are two more scalings of the phase diagrams not shown in the main text.
All parameters may be scaled by $\sqrt u \bar \varphi_1$ or $\alpha$, in which case the correlation lengths take the form (with prime indicating rescaled parameters)
\begin{align}
  \kappa_\pm' &= r' + 3\left(1 + {v'}_1^2\right) \pm \sqrt{9\left(1 - {v'}_1^2\right)^2 - \alpha'^2 },&
  \kappa_\pm'' &= r'' + 3\left({v''}_1^2 + {v''}_2^2\right) \pm \sqrt{9\left({v''}_1^2 - {v''}_2^2\right)^2 - 1},
\end{align}
correspondingly.
These scalings give two new 3D phase diagrams, and 
two new sets of 2D slices of these 3D shapes.
These are shown in \autoref{fig: more C4 diagrams}.
In \autoref{fig:eigen}, we illustrate the eigenvalue $\lambda_+(\bm q; \bm k)$ responsible for the Eckhaus-instability of the traveling wave solution.
The two top values are perturbation parallel to the direction and travel, the real part on top and imaginary below.
The two bottom plots are perturbations orthogonal to this, again, the real part on top and imaginary below.

\section{Videos}

There are three videos included as supplementary material, referenced in the main text.
\autoref{table} gives an overview of the videos, and their parameter values.
All simulations used a time step of $\Delta t = 1.6 \cdot 10^{-5}$.

\begin{table}[h]
    \setlength{\tabcolsep}{12pt}
	\caption{
		The videos included as supplemental material, and the parameter values used to generate them.
		}
	\begin{tabular}{ c c c c c c} 
	\hline
	Video nr. & $u,-r$ & $\alpha$ & $D$ & $\bar \varphi_1$ & $\bar \varphi_2$  \\
	\hline
    1 & 10 & 2  & $2\times10^{-4}$ & -0.2  & -0.1 \\ 
    2 & 10 & 1  & $2\times10^{-4}$ & -0.1  & 0    \\
	3 & 40 & 0  & $2\times10^{-4}$ & -0.8  & 0     \\
	\hline
	\end{tabular}
	\label{table}
\end{table}

